\documentclass[
	preprint,
	floats,floatfix,superscriptaddress,
	preprintnumbers,showpacs,nofootinbib
]{revtex4-1}
\usepackage{
	amsfonts,amsmath,amssymb,latexsym,
	eucal,array,subfigure,bm,mathrsfs,
	ulem,epsfig,physics,color,graphicx,
	mathtools
	}

\begin{document}
\preprint{RUP-26-13}
%---------------------------------------------------------------------%

%---------------------------------------------------------------------%
\title{Determining Kerr black hole spin and inclination from a segment of the critical curve in black hole images}
%---------------------------------------------------------------------%
%---------------------------------------------------------------------%
\author{Kenta Hioki}
\email{kenta.hioki@gmail.com}
\noaffiliation
\author{Umpei Miyamoto}
\email{umpei@akita-pu.ac.jp}
\affiliation{Research and Education Center for Comprehensive Science, Akita Prefectural University, Akita 015-0055, Japan}
\author{Tomohiro Harada}
\email{harada@rikkyo.ac.jp}
\affiliation{Department of Physics, Rikkyo University, Toshima, Tokyo 171-8501, Japan}
%\date{\today}
%---------------------------------------------------------------------%
%---------------------------------------------------------------------%
\begin{abstract}
We present a method for determining the spin parameter $a/M$ and
inclination angle $i$ of a non-extremal Kerr black hole from
segments of the critical curve identified in black hole images.
Although the critical curve itself is not directly observable,
higher-order photon rings accumulate near it,
and in realistic observations localized portions of the resulting brightness
enhancement may be available for identifying segments of the
critical curve.
We introduce standardized segments of the critical curve and define three observables that characterize their geometry.
We show that these observables uniquely determine $(a/M,i)$, together
with an auxiliary parameter $r_{nl}\in[0,1]$ specifying the location
of the identified segment along the critical curve, within the domain
considered.
Thus, even a segment of the critical curve contains sufficient geometric
information to constrain the black hole spin and inclination without
reconstructing the full critical curve.
The framework is naturally suited to realistic observations and may be
extended to more general rotating black hole spacetimes.
\end{abstract}
%---------------------------------------------------------------------%
%---------------------------------------------------------------------%
\maketitle
%\tableofcontents

%-------------------------------------------%
\section{Introduction}
\label{sec:intro}
%-------------------------------------------%
Horizon-scale imaging observations of supermassive black holes
by the Event Horizon Telescope (EHT)
have opened a new observational window into the strong-gravity regime~\cite{EventHorizonTelescope:2019dse, EventHorizonTelescope:2022wkp}.
The observed images exhibit a bright ring-like structure
surrounding a central brightness depression,
commonly associated with the photon ring
and the underlying critical curve of null geodesics
\cite{Bardeen:1973xx, Chandrasekhar:1985kt}.
These features encode information about the spacetime geometry
and thus provide a promising avenue
for determining black hole parameters.

A number of studies have proposed methods
to extract physical parameters from black hole images,
typically based on characteristics of the shadow,
such as its size and shape~\cite{Hioki:2008zw, Hioki:2009na, Tsukamoto:2014tja, Abdujabbarov:2015xqa, EventHorizonTelescope:2021dqv, Hioki:2022mdg, Hioki:2023ozd, Hioki:2024vta}.
Such approaches are effective when the full contour of the shadow is accurately resolved.
However, current observations are limited by finite angular resolution, sparse interferometric sampling, and astrophysical variability~\cite{EventHorizonTelescope:2019dse, EventHorizonTelescope:2019ths, EventHorizonTelescope:2022wkp}.
As a consequence, the full contour of the shadow cannot be reliably identified.

These observational limitations imply that methods relying on the full contour of the shadow are not directly applicable to current observations. Instead, one may seek to extract information from partial image features. This raises the question of whether black hole parameters can be determined from partial information contained in the image.

The formation of the observed ring-like structure depends not only on the geodesic geometry
but also on the properties of the emitting accretion flow.
EHT observations are well described by radiatively inefficient accretion flows,
which are geometrically thick, optically thin,
and exhibit smoothly varying density and temperature profiles~\cite{Narayan:1994xi, Yuan:2014gma, EventHorizonTelescope:2019pgp, EventHorizonTelescope:2019ths}.
Time-dependent three-dimensional general relativistic magnetohydrodynamic simulations further indicate that the emissivity,
magnetic field, and plasma properties vary smoothly
over radial scales of several gravitational radii~\cite{Dexter:2016cdk, EventHorizonTelescope:2019pgp, EventHorizonTelescope:2019ths}.

These results suggest that, although the emissivity and transfer effects are not uniform,
they typically vary on scales larger than the photon orbit scale.
Under such conditions,
the geometric accumulation of photon trajectories near unstable spherical photon orbits
dominates the formation of the brightness profile,
producing a narrow brightness structure that traces the critical curve.
In general, brightness enhancement does not coincide exactly with the critical curve,
due to radiative effects and finite observational resolution.
Nevertheless, when the emissivity and transfer functions
vary sufficiently smoothly so that they do not dominate
over the geometric accumulation,
the brightness enhancement follows the shape of the critical curve.

The narrow brightness structure manifests itself
not as a set of individually resolved photon rings,
but as a continuous brightness enhancement
in the intensity distribution~\cite{Gralla:2019xty}.
We therefore define the observable feature intrinsically from the image
as the locus of points at which the intensity exhibits
a non-degenerate local maximum in a transverse direction.
More precisely,
these points are characterized by the condition that
the directional derivative of the intensity vanishes along a certain direction,
and the second derivative along that direction is negative,
while variation in the orthogonal direction is allowed.
We refer to this set of points as the ridge of the intensity.
This ridge corresponds to this brightness enhancement, often referred to as the lensing ring.

In this work, we therefore assume that localized segments of the ridge
can be identified from observational data.
Under this assumption,
we focus on segments of the critical curve,
namely, connected subsets of the critical curve
identified via the ridge of the intensity.
In particular, the ridge serves as an observational proxy
for the critical curve at the level of local geometry,
since it traces the accumulation of photon trajectories
that asymptotically approach it.

In this paper, we introduce a systematic framework for analyzing standardized segments of the critical curve, in which the distance between the endpoints of each segment is fixed. We show that the mapping from the black hole parameters to the corresponding standardized segments of the critical curve is injective. In particular, we demonstrate that even a segment of the critical curve, identified via the intrinsic ridge of the intensity, encodes sufficient information to uniquely determine the dimensionless parameters of a Kerr black hole.

Looking ahead, future interferometric missions,
such as the Black Hole Explorer~\cite{Johnson:2019ljv, Akiyama:2024msp, Johnson:2024ttr, Lupsasca:2024xhq},
aim to achieve substantially higher angular resolution
by extending very-long-baseline interferometry to space.
These observations are expected to probe the fine structure
of the photon ring and the critical curve with unprecedented precision.
Nevertheless, robust parameter inference will continue to rely on methods
that can operate on partial and localized features,
owing to noise, variability, and incomplete reconstruction.
The present framework is therefore well suited
both for current observations and for next-generation experiments.

The structure of this paper is as follows.
In Sec.~\ref{sec:set}, we review null geodesic motion in Kerr spacetime
and specify the observational setup, and describe how segments of the critical curve arise from null geodesics associated with
unstable spherical photon orbits.
In Sec.~\ref{sec:spo}, we analyze spherical photon orbits and clarify
their role in the formation of the critical curve.
In Sec.~\ref{sec:pho}, we introduce segments of the critical curve as
connected subsets of the critical curve on the observer's screen and
discuss their properties for Kerr black holes.
In Sec.~\ref{sec:sopccf}, we present a standardization procedure for
segments of the critical curve and define standardized segments of the critical curve, showing that they form a one-parameter family of embedded curve segments and describing
their parametrization.
In Sec.~\ref{sec:oc}, we construct observables that characterize
standardized segments of the critical curve by introducing a Fourier representation of the curve shape and
defining principal-component observables that capture its essential
geometric features.
In Sec.~\ref{sec:pm}, we demonstrate that these observables enable the
determination of the parameters
$(a,i,r_{nl})$ by establishing the injectivity of the mapping
from parameters to observables.
In Sec.~\ref{sec:demo}, we evaluate the numerical accuracy of the parameter-determination method.
Finally, we summarize our results and discuss future prospects.

Throughout this paper, we use geometrized units with $G=c=1$,
so that the black hole mass $M$ defines the natural length scale.
All quantities with dimensions of length are expressed in units of $M$,
and we identify each such quantity with its dimensionless counterpart,
thereby simplifying notation and focusing on the underlying geometry.

%-------------------------------------------%
\section{Setup}
\label{sec:set}
%-------------------------------------------%
%-------------------------------------------%
\subsection{Geodesic motion in Kerr spacetime}
%-------------------------------------------%
The spacetime of a rotating black hole is widely believed to be well
described by the Kerr metric~\cite{Kerr}.
In Boyer--Lindquist coordinates
$x^\mu=(t,r,\theta,\phi)$ $(\mu,\nu=0,1,2,3)$,
the line element can be written as
\begin{eqnarray}
	g_{\mu \nu}\,{\rm d}x^\mu {\rm d}x^\nu 
	&=&
	-\left(
		1-\frac{2r}{\varSigma}
	\right) {\rm d}t^2
	+
	\frac{\varSigma}{\varDelta} {\rm d}r^2
	+
	\varSigma {\rm d}\theta ^2
	-\frac{4ra\sin^2\theta}{\varSigma} {\rm d}t {\rm d}\phi
	+
	\frac{A \sin ^2 \theta}{\varSigma} {\rm d}\phi ^2 ,
	\label{eq:metric}
\end{eqnarray}
where
\begin{eqnarray}
    \varSigma
	&:=&
	r^2 + a^2 \cos^2 \theta, \\
    \varDelta
	&:=&
	r^2 - 2r + a^2, \\
    A
	&:=&
	\left( r^2 + a^2 \right)^2 - a^2 \varDelta \sin^2\theta .
\end{eqnarray}
As stated in Sec.~\ref{sec:intro}, we identify each length quantity with its
dimensionless counterpart.
Accordingly, $r$, $t$, and $a$ are dimensionless,
while $\theta$ and $\phi$, being angular coordinates, are dimensionless by definition.
Here $a$ denotes the dimensionless spin parameter.
If $|a|\le 1$, the spacetime possesses an event horizon and describes a black hole.
The radii of the outer and inner horizons are denoted by $r_{+}$ and $r_{-}$, respectively.

To determine the trajectory $x^\mu(\lambda)$ of a massless test particle,
interpreted as a light ray, we consider the geodesic equation, where
$\lambda$ is an affine parameter along the null geodesic.
Because Kerr spacetime admits four constants of motion in involution,
the geodesic motion is completely integrable.

We begin with the Lagrangian
\begin{eqnarray}
	\mathcal{L}
	=
	\frac{1}{2} g_{\mu\nu}\dot{x}^\mu \dot{x}^\nu ,
	\qquad
	\dot{x}^\mu \coloneqq \frac{{\rm d}x^\mu}{{\rm d}\lambda},
	\label{eq:lag}
\end{eqnarray}
from which the conserved energy $E$ and the axial component of the
angular momentum $L$ are obtained as
\begin{eqnarray}
	E
	&\coloneqq&
	- \frac{\partial \mathcal{L}}{\partial \dot{t}}
	=
	\left(1-\frac{2r}{\varSigma}\right)\dot{t}
	+
	\frac{2ra\sin^2\theta}{\varSigma}\dot{\phi}, \\
	L
	&\coloneqq&
	\frac{\partial \mathcal{L}}{\partial \dot{\phi}}
	=
	-\frac{2ra\sin^2\theta}{\varSigma}\dot{t}
	+
	\frac{A\sin^2\theta}{\varSigma}\dot{\phi}.
	\label{eq:angmom}
\end{eqnarray}
Together with the Lagrangian $\mathcal{L}$, these quantities are
constants of motion.
In addition, Kerr spacetime admits another conserved quantity
$\mathcal{Q}$, known as the Carter constant~\cite{Chandrasekhar:1985kt}.

For null geodesics ($\mathcal{L}=0$), it is convenient to introduce the
dimensionless conserved quantities
\begin{eqnarray}
	\xi \coloneqq \frac{L}{E}, \qquad
	\eta \coloneqq \frac{\mathcal{Q}}{E^2}.
\end{eqnarray}
We also define the four-momentum of a massless test particle by
\begin{eqnarray}
	k^\mu \coloneqq \frac{{\rm d}x^\mu}{{\rm d}\tilde{\lambda}} ,
\end{eqnarray}
where the rescaled affine parameter $\tilde{\lambda}$ is defined by
$\tilde{\lambda} \coloneqq E\lambda$.
With these definitions, the equations of motion can be written in a form independent of $E$ as
\begin{eqnarray}
	\varSigma k^t
	&=&
	\frac{A-2ra\xi}{\varDelta},
	\label{eq:velocity2}
	\\
	\varSigma k^r
	&=&
	\pm \sqrt{R},
	\label{eq:velocity0}
	\\
	\varSigma k^\theta
	&=&
	\pm \sqrt{\varTheta},
	\label{eq:velocity1}
	\\
	\varSigma k^\phi
	&=&
	\frac{2ra+\xi\csc^2\theta(\varSigma-2r)}{\varDelta},
	\label{eq:velocity3}
\end{eqnarray}
where
\begin{eqnarray}
	K
	&\coloneqq&
	\eta + (a-\xi)^2 ,
	\label{eq:cqk}
	\\
	R(r)
	&\coloneqq&
	\left(r^2+a^2-a\xi\right)^2 - K\varDelta ,
	\\
	\varTheta(\theta)
	&\coloneqq&
	K - (a\sin\theta-\xi\csc\theta)^2 .
\end{eqnarray}

%-------------------------------------------%
\subsection{Observational setup and emission model}
%-------------------------------------------%
Next, we specify the observational setup for an observer detecting
null rays.
Since we consider distant black holes as observational targets,
we assume a static observer, as is standard in black hole
imaging studies.
Possible observer frames include zero-angular-momentum observers
(ZAMOs)~\cite{Bardeen:1973xx} and Carter observers~\cite{Carter:1968rr},
which exhibit different azimuthal motions~\cite{Chang:2020lmg};
however, at sufficiently large distances these frames coincide and
reduce to the same static observer.
We therefore adopt the ZAMO frame.

We assume that the observer is located at a radial distance $r_o$
from the black hole with inclination angle $i$.
Since Eq.~(\ref{eq:metric}) is defined for $\theta \in (0,\pi)$,
we restrict the inclination to
\begin{eqnarray}
	i \in (0,\pi/2],
\end{eqnarray}
thereby excluding observers on the rotation axis.
For $i=0$, the azimuthal direction becomes degenerate, and a different
coordinate description would be required.

With this setup, we introduce the orthonormal tetrad
\begin{eqnarray}
	e_{(t)}
	&\coloneqq&
	\sqrt{\frac{A}{\varSigma\varDelta}}
	\left(
	\partial_t
	+
	\frac{2ar}{A}\partial_\phi
	\right),
	\\
	e_{(r)}
	&\coloneqq&
	-
	\sqrt{\frac{\varDelta}{\varSigma}}
	\partial_r,
	\\
	e_{(\theta)}
	&\coloneqq&
	\frac{1}{\sqrt{\varSigma}}
	\partial_\theta,
	\\
	e_{(\phi)}
	&\coloneqq&
	-
	\sqrt{\frac{\varSigma}{A}}
	\csc\theta\,\partial_\phi .
	\label{eq:tetrad}
\end{eqnarray}
The timelike vector
$e_{(t)}|_{(r,\theta)=(r_o,i)}$
represents the four-velocity of the observer,
while the spacelike vector
$e_{(r)}|_{(r,\theta)=(r_o,i)}$
points toward the black hole.

We assume that the emitting source is localized and spatially
non-uniform.
A source is said to be localized if its emission originates from a
finite region, rather than
from a spatially uniform background extending to infinity.
It is non-uniform if the emissivity varies within the emitting region.
In addition, we assume that the emissivity varies smoothly
on spatial scales comparable to or larger than the photon orbit scale.
These assumptions are satisfied by physically relevant models of
black hole environments; in particular, radiatively inefficient
accretion flows predict millimeter emission from a geometrically
thick, optically thin plasma with spatially varying density,
temperature, and magnetic field~\cite{Narayan:1994xi, Yuan:2014gma}.
General relativistic magnetohydrodynamic simulations used in
Event Horizon Telescope imaging likewise indicate that the radiation
originates from a confined and non-uniform emitting region
and varies smoothly over radial scales of several gravitational radii
\cite{EventHorizonTelescope:2019pgp, EventHorizonTelescope:2019ths}.

Under such conditions, strong gravitational lensing near the black hole concentrates radiation along families
of nearly bound null geodesics, thereby producing photon-ring
features in black hole images~\cite{Gralla:2019xty}.
Because the emission is localized and non-uniform,
the observed intensity is typically enhanced only over restricted
portions of these geodesic families.
As a result, instead of a complete and uniformly bright photon ring,
one naturally obtains localized segments of enhanced brightness,
which appear as ridge-like features in the image.
These features trace the critical curve in the sense that
they arise from the geometric accumulation of photon trajectories,
while their partial visibility reflects the underlying
non-uniformity of the emission.

This provides a physical basis for the appearance of
the segment of the critical curve,
which we define as the connected subset of the critical curve
identified through the ridge of the observed intensity distribution.

%-------------------------------------------%
\section{Spherical photon orbits}
\label{sec:spo}
%-------------------------------------------%
%-------------------------------------------%
\subsection{Conserved quantities}
%-------------------------------------------%
We focus on spherical photon orbits, which form an important class
of null geodesics relevant to photon rings.
A null geodesic with constant radial coordinate is referred to as
a spherical photon orbit.

The radius $r_s$ of a spherical photon orbit is determined by the
conditions~\cite{deVries:1999tiy}
\begin{eqnarray}
	R(r_s) = 0 , \qquad
	\frac{{\rm d}R}{{\rm d}r}(r_s) = 0 ,
	\label{eq:spo}
\end{eqnarray}
and the polar motion must satisfy
\begin{eqnarray}
	\varTheta(\theta) \ge 0 ,
	\label{eq:contheta}
\end{eqnarray}
for a spherical photon orbit of radius $r_s$ to exist.

The conserved quantities $\xi$ and $\eta$ satisfying
Eq.~\eqref{eq:spo} are denoted by $\xi_s$ and $\eta_s$,
respectively, and are given by~\cite{deVries:1999tiy}
\begin{eqnarray}
	\xi_s
	&=&
	\frac{r_s^2+a^2}{a}
	-
	\frac{2r_s\varDelta}{a(r_s-1)} ,
	\label{eq:spoell}
	\\
	\eta_s
	&=&
	-
	\frac{r_s^3\left[r_s(r_s-3)^2-4a^2\right]}
	{a^2(r_s-1)^2} .
	\label{eq:spoq}
\end{eqnarray}
These quantities are conserved along spherical photon orbits
with radius $r_s$.
Taking into account the position of the observer,
Eqs.~(\ref{eq:contheta}), (\ref{eq:spoell}), and (\ref{eq:spoq})
imply that the radius $r_s$ must satisfy
\begin{eqnarray}
	\varTheta(i)
	\Big|_{(\xi,\eta)=(\xi_s,\eta_s)}
	\ge 0 .
\end{eqnarray}

Substituting Eqs.~(\ref{eq:spoell}) and (\ref{eq:spoq})
into Eq.~(\ref{eq:cqk}), we obtain
\begin{eqnarray}
	K_s \coloneqq
	\frac{4r_s^2\varDelta}{(r_s-1)^2} .
	\label{eq:cartercnst}
\end{eqnarray}
Since $K$ is non-negative in general~\cite{Chandrasekhar:1985kt} and since we consider only spherical photon orbits outside the event horizon,
the radius $r_s$ must lie in
\begin{eqnarray}
	r_s \in (r_+,\infty) .
	\label{eq:spoexist}
\end{eqnarray}

%-------------------------------------------%
\subsection{Radial instability}
%-------------------------------------------%
The condition for radial instability of a spherical photon orbit is
\begin{eqnarray}
	\frac{{\rm d}^2R}{{\rm d}r^2}(r_s) > 0 .
	\label{eq:sta}
\end{eqnarray}
The function ${\rm d}^2R/{\rm d}r^2$ has at most two real roots,
namely $0$ and $r_u$, where
\begin{eqnarray}
	r_u
	\coloneqq
	1-\left(1-a^2\right)^{1/3}
\end{eqnarray}
denotes the non-negative real root.
Accordingly, Eq.~(\ref{eq:sta}) is equivalent to
\begin{eqnarray}
	r_s \in (r_u,\infty) .
	\label{eq:spounsta}
\end{eqnarray}

Combining Eqs.~(\ref{eq:spoexist}) and (\ref{eq:spounsta}),
we obtain
\begin{eqnarray}
	r_s \in (r_+,\infty) .
	\label{eq:rsta}
\end{eqnarray}
Therefore, all spherical photon orbits outside the event horizon
that are relevant for photon-ring formation are radially unstable.

%-------------------------------------------%
\section{Segments of the critical curve indicated in black hole images}
\label{sec:pho}
%-------------------------------------------%
\begin{figure}[tb]
	\begin{tabular}{ cccc }
		\includegraphics[height=3.7cm]{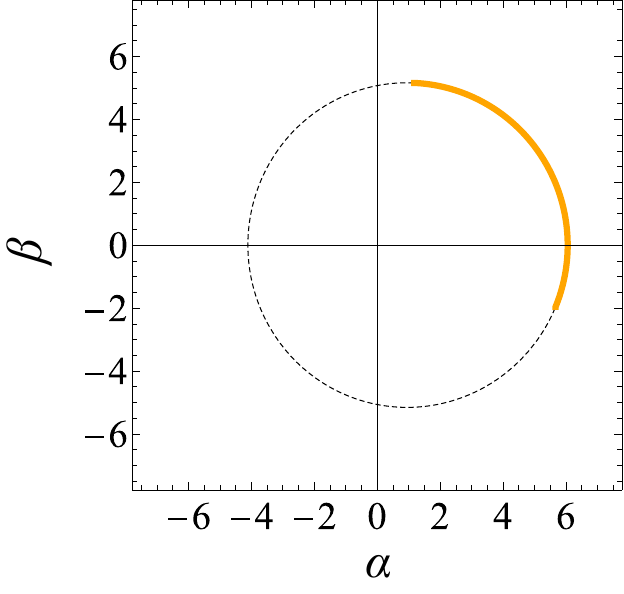} &
		\includegraphics[height=3.7cm]{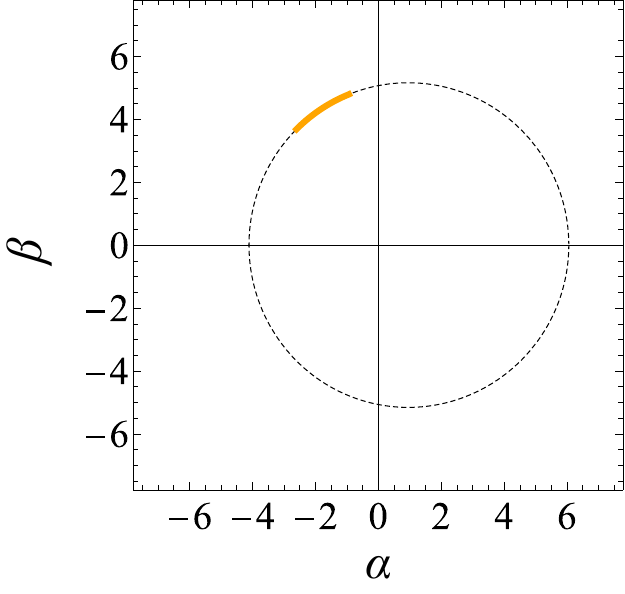} &
		\includegraphics[height=3.7cm]{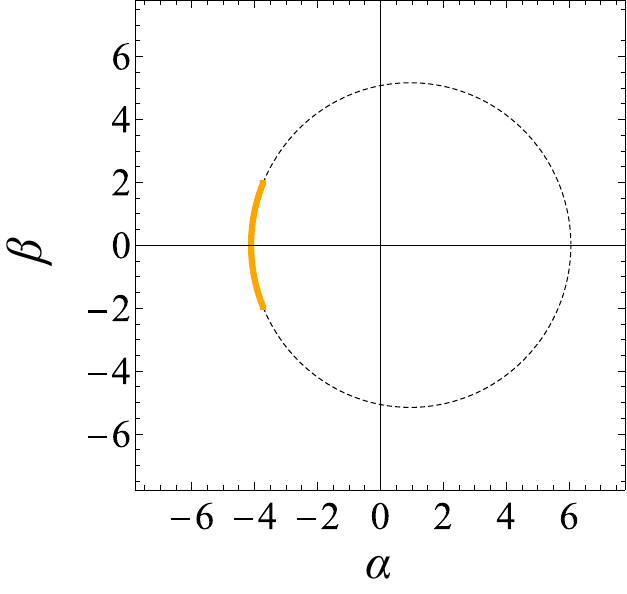} &
		\includegraphics[height=3.7cm]{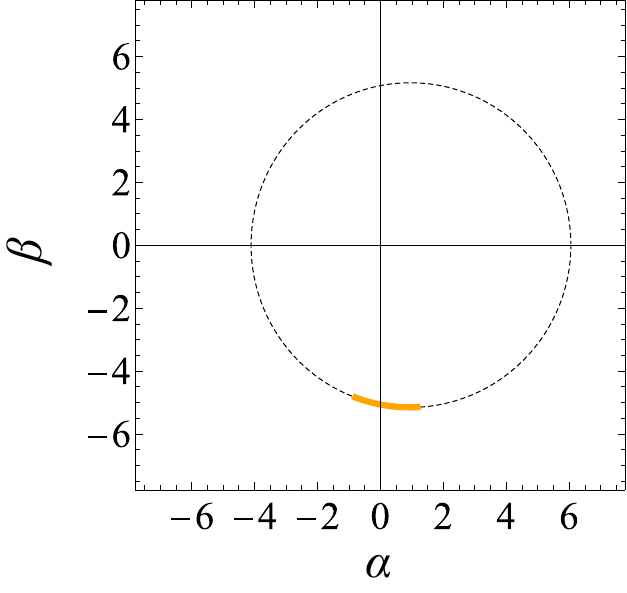} \\
		(a) & (b) & (c) & (d) \\
	\end{tabular}
	\caption{\footnotesize{Examples of segments of the critical curve extracted from a single critical curve $C(a,i)$ of a non-extremal Kerr black hole with $(a,i)=(0.590,0.927)$.
In each panel, the extracted segment of the critical curve is shown by the orange solid line, while the full critical curve is shown by the orange solid line together with the black dashed line.
In panels (a)--(d), the interval pairs $\left([\epsilon_1, \epsilon_1 + \delta_1], [\epsilon_2, \epsilon_2 + \delta_2]\right)$ are given by
$\left([2.97, 3.46], [3.42, 3.46]\right)$,
$\left([2.51, 2.72], \emptyset\right)$,
$\left([2.31, 2.36], [2.31, 2.36]\right)$,
and
$\left(\emptyset, [2.74, 2.97]\right)$, respectively.}}
	\label{fig010}
\end{figure}
The observer's screen is defined as the two-dimensional plane
perpendicular to the line-of-sight direction $e_{(r)}$.
The Bardeen coordinates $(\alpha,\beta)$ on this screen are defined by
\begin{eqnarray}
\alpha
&\coloneqq&
\lim_{r_o \to \infty}
\frac{-r_o k^{(\phi)}}{k^{(t)}}
=
- \xi \csc i ,
\label{eq:alpha}
\\
\beta
&\coloneqq&
\lim_{r_o \to \infty}
\frac{r_o k^{(\theta)}}{k^{(t)}}
=
\left(
\eta
+
a^2 \cos^2 i
-
\xi^2 \cot^2 i
\right)^{1/2}.
\label{eq:beta}
\end{eqnarray}
Here $\left(k^{(t)},k^{(r)},k^{(\theta)},k^{(\phi)}\right)$ denote the
tetrad components of the photon four-momentum measured by the observer
at infinity.
The coordinates $(\alpha,\beta)$ therefore provide Cartesian
coordinates on the observer's image plane.

All quantities with dimensions of length are expressed in units of the black hole mass $M$.
Accordingly, the coordinates $(\alpha, \beta)$ and parameter $r_s$ are treated as dimensionless quantities.

The point on the screen corresponding to a null geodesic
asymptotically approaching an unstable spherical photon orbit
of radius $r_s$ is denoted by
\begin{eqnarray}
\gamma_\pm(r_s;a,i)
\coloneqq
\bigl(
\alpha(r_s;a,i),
\pm \beta(r_s;a,i)
\bigr).
\label{eq:defgammas}
\end{eqnarray}
Here, $\alpha(r_s;a,i)$ and $\beta(r_s;a,i)$ are obtained by evaluating Eqs.~\eqref{eq:alpha} and \eqref{eq:beta} at $\xi=\xi_s(r_s)$ and $\eta=\eta_s(r_s)$, where $\xi_s$ and $\eta_s$ are given by Eqs.~\eqref{eq:spoell} and \eqref{eq:spoq}.

We first consider the non-extremal Kerr black hole case
$0 < a < 1$.
The critical curve $C(a,i)$ on the observer's screen is given by
\begin{eqnarray}
C(a,i)
&=&
\left\{
\gamma_+(r_s;a,i)
\,\middle|\,
r_s\in I(a,i)
\right\}
\nonumber
\\
&&
\cup
\left\{
\gamma_-(r_s;a,i)
\,\middle|\,
r_s\in I(a,i)
\right\},
\end{eqnarray}
where $I(a,i)$ denotes the set of radii corresponding to
unstable spherical photon orbits for the parameters $(a,i)$.

A segment of the critical curve is defined as a connected subset of $C(a,i)$.
Using two intervals
$[\epsilon_1,\epsilon_1+\delta_1]$
and
$[\epsilon_2,\epsilon_2+\delta_2]$
contained in $I(a,i)$,
the segment of the critical curve can be written as
\begin{eqnarray}
&&\left\{
\gamma_+(r_s;a,i)
\,\middle|\,
r_s\in[\epsilon_1,\epsilon_1+\delta_1]
\right\}
\nonumber
\\
&&
\cup
\left\{
\gamma_-(r_s;a,i)
\,\middle|\,
r_s\in[\epsilon_2,\epsilon_2+\delta_2]
\right\}.
\end{eqnarray}
The intervals are chosen so that the segment of the critical curve is connected, and either
interval may be empty.
In actual observations, multiple disconnected segments of the critical curve may be detected; in such cases, we analyze a single connected segment.
Examples of segments of the critical curve are shown in
Fig.~\ref{fig010}.

For comparison, in the Schwarzschild case $a = 0$, the critical
curve is the circle
\begin{eqnarray}
\alpha^2 + \beta^2 = 27 ,
\end{eqnarray}
centered at the origin with radius $3\sqrt{3}$~\cite{Synge:1966okc}.
Any segment of the critical curve is therefore an arc of this circle.

In the extremal Kerr case $a=1$, the structure of the critical
curve differs qualitatively from that in the non-extremal case.
In particular, the set of spherical photon orbits admits sequences
whose radii $r_s$ approach the horizon radius $r_{+}=1$,
so that $r_{+}$ becomes an accumulation point.
The images of these near-horizon photon orbits on the observer's
screen collapse onto a vertical segment known as the Near-Horizon Extreme Kerr (NHEK) line~\cite{Gralla:2017ufe}.
Consequently, the parametrization of the critical curve by
$r_s$ becomes degenerate in the limit $r_s\to r_{+}$,
and the curve consists of a regular part together with the NHEK-line
segment.
Thus, unlike the non-extremal case, the extremal Kerr critical curve
cannot be described solely by the $r_s$-parametrized branches
$\gamma_\pm$.

The presence of the NHEK line implies that, for an extremal Kerr black
hole, segments of the critical curve may reduce to straight line
segments and hence are not unique with respect to the black hole
parameters $(a,i)$.
Accordingly, in this paper we restrict attention to non-extremal Kerr
black holes with $0 < a < 1$.

%-------------------------------------------%
\section{Standardization of Segments of the Critical Curve}
\label{sec:sopccf}
%-------------------------------------------%
\begin{figure}[tb]
	\begin{tabular}{ c }
		\includegraphics[height=4.4cm]{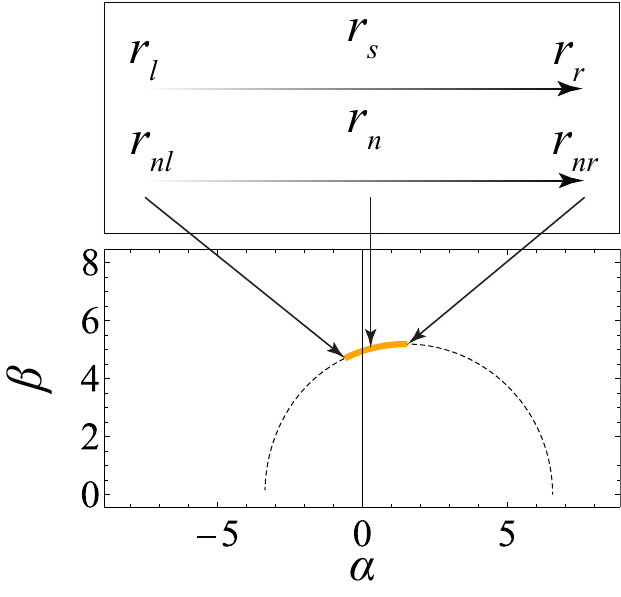}
	\end{tabular}
	\caption{\footnotesize{A standardized segment of the critical curve parametrized by $r_s$
(or equivalently $r_n$) for fixed black hole parameters $(a,i)$
(orange solid line).
The values $r_s=r_l$ ($r_n=r_{nl}$) and $r_s=r_r$ ($r_n=r_{nr}$)
correspond to the left and right endpoints of the segment, respectively.}}
	\label{fig025}
\end{figure}
\begin{figure}[tb]
	\begin{tabular}{ c }
		\includegraphics[height=4.4cm]{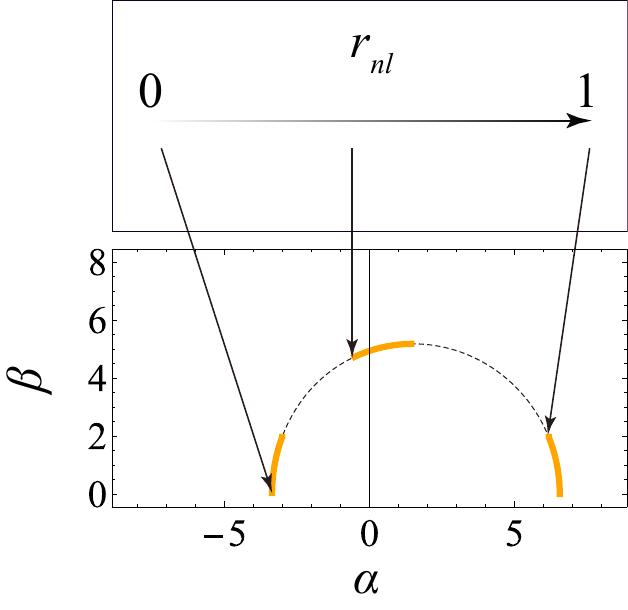}
	\end{tabular}
	\caption{\footnotesize{Standardized segments of the critical curve for fixed black hole parameters $(a,i)$ (orange solid lines).
These segments form a one-parameter family parametrized by $r_{nl}$.
The value $r_{nl}=0$ corresponds to the segment whose
left endpoint lies on the horizontal axis, whereas $r_{nl}=1$ corresponds to the segment whose right endpoint lies on the horizontal axis.}}
	\label{fig030}
\end{figure}
We present a concrete method for determining black hole
parameters from segments of the critical curve.
To this end, we introduce a procedure for extracting
standardized segments of the critical curve from identified segments of the critical curve.

As the distance between the endpoints of a segment of the critical curve decreases, its shape approaches that of a short straight line segment.
It is therefore difficult to distinguish among segments of the critical curve whose endpoints are separated by only a very small distance.
We assume that a segment of the critical curve whose endpoints
are separated by a distance of at least $2$ has been identified.
Here, the distance refers to the Euclidean distance between the endpoints on the screen, not to the length of the segment itself.

Such segments may intersect the $\alpha$ axis;
however, we assume that the portion whose endpoints are separated
by a distance of at least $2$ lies entirely either in the upper
or in the lower half-plane of the screen.
If this portion lies in the lower half-plane, we reflect it across
the $\alpha$ axis.
It therefore suffices to analyze features contained in the
upper half-plane.

We consider the parameter region $P$ defined by
\begin{eqnarray}
P \coloneqq
\left\{
(a,i)
\,\middle|\,
0 < a < 1,\;
0 < i \le \frac{\pi}{2}
\right\}.
\label{eq:pspace}
\end{eqnarray}

From the identified segments of the critical curve we extract
segments whose endpoints are separated by a distance of exactly $2$.
We refer to such segments as standardized segments of the critical curve.
Although the argument remains valid for segments whose endpoints are separated by a distance of less than $2$, it is convenient for practical numerical analysis to choose segments whose endpoints are separated by a distance of $2$.

Let $r_l$ and $r_r$ denote the values of $r_s$ corresponding to the left and
right endpoints of the standardized segment of the critical curve for the parameters $(a,i)$.
These radii satisfy
\begin{eqnarray}
\|
\gamma_{+}(r_l; a ,i)
-
\gamma_{+}(r_r; a ,i)
\|
=2 ,
\label{eq:dist2}
\end{eqnarray}
where $\|\cdot\|$ denotes the Euclidean norm on the observer's screen.
For a given $r_l$, the value $r_r=r_r(r_l)$ is uniquely determined
by Eq.~\eqref{eq:dist2}.
Indeed, for fixed $r_l$, the left-hand side of Eq.~\eqref{eq:dist2}
is a strictly increasing function of $r_r$.

As illustrated in Fig.~\ref{fig025}, the standardized segment of the critical curve can therefore be written as
\begin{eqnarray}
\left\{
\gamma_+(r_s;a,i)
\,\middle|\,
r_s\in[r_l, r_r]
\right\},
\end{eqnarray}
where
\begin{eqnarray}
[r_l, r_r]
\subset
I(a,i).
\end{eqnarray}
We now reparametrize the standardized segment of the critical curve by introducing a normalized parameter
$r_n \in [r_{nl}, r_{nr}] \subset [0,1]$, where $r_n=r_{nl}$ and
$r_n=r_{nr}$ correspond to the left and right endpoints, respectively.
For fixed black hole parameters $(a,i)$, the standardized segment of the critical curve is uniquely
specified by the left endpoint $r_{nl}\in[0,1]$.
The details of this reparametrization are discussed in
Appendix~\ref{sec:re}.

As shown in Fig.~\ref{fig030}, for fixed parameters $(a,i)$,
the standardized segments of the critical curve form a one-parameter family with respect to $r_l$,
or equivalently $r_{nl}$.

%-------------------------------------------%
\section{Observables Characterizing Segments of the Critical Curve}
\label{sec:oc}
%-------------------------------------------%
%-------------------------------------------%
\subsection{Fourier representation of standardized segments of the critical curve}
%-------------------------------------------%
\begin{figure}[tb]
	\begin{tabular}{ c }
		\includegraphics[height=4.4cm]{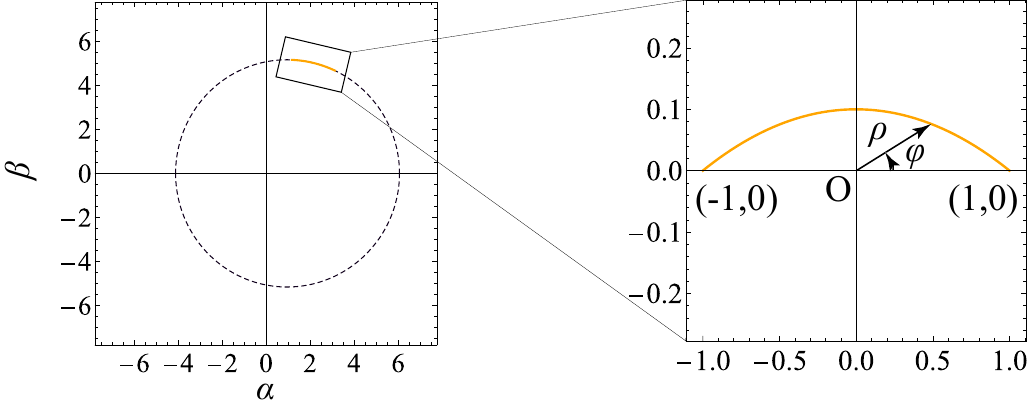}
	\end{tabular}
	\caption{\footnotesize{Standardized segment of the critical curve (orange solid line) and polar coordinates $(\rho,\varphi)$. A segment whose endpoints are separated by a distance of $2$ is selected from the critical curve, shown by the orange solid line together with the black dashed line, and defined as a standardized segment of the critical curve. Here, the distance refers to the Euclidean distance between the endpoints on the screen, not to the length of the segment itself. The screen coordinates are then rotated and translated so that the two endpoints of the segment are mapped to $(-1,0)$ and $(1,0)$. Polar coordinates $(\rho,\varphi)$ are introduced in the resulting coordinate system.}}
	\label{fig020}
\end{figure}
\begin{figure}[tb]
	\begin{tabular}{ccc}
		\includegraphics[height=4.6cm]{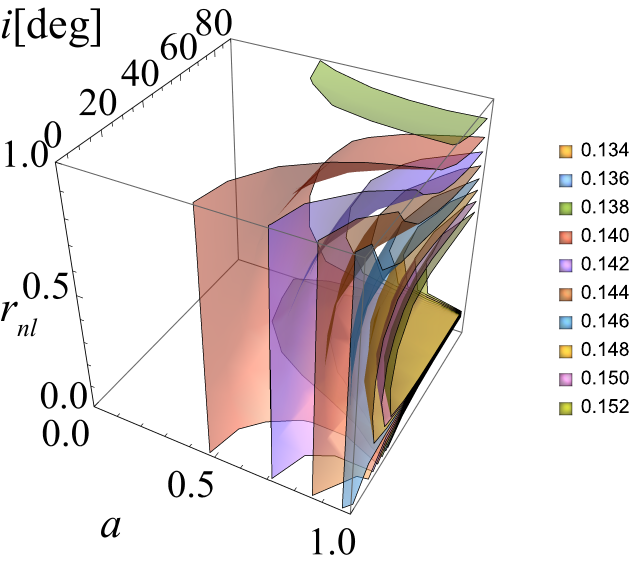} &
		\includegraphics[height=4.6cm]{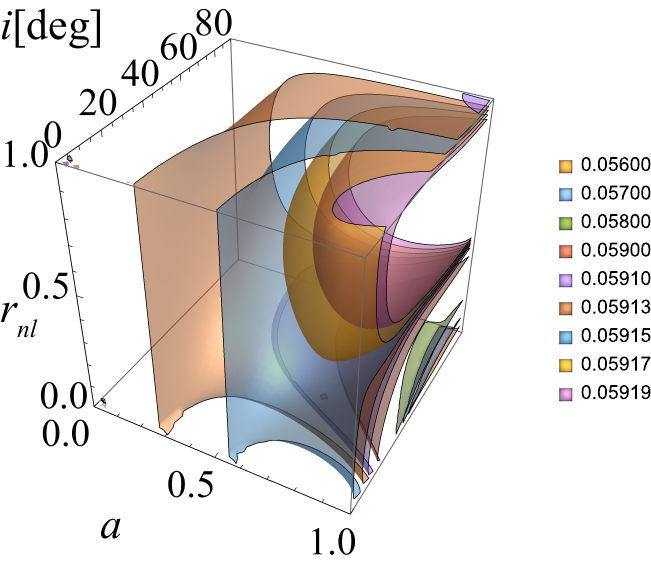} &
		\includegraphics[height=4.6cm]{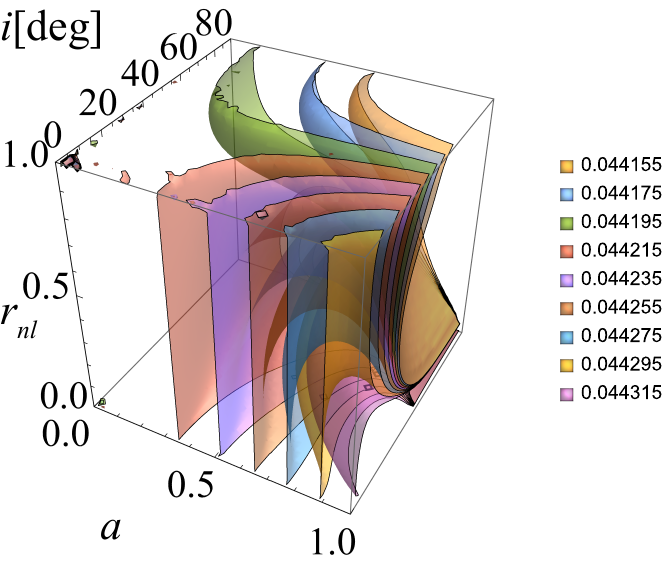} \\
		(a) $z_1$ & (b) $z_2$ & (c) $z_3$ \\
	\end{tabular}
	\caption{\footnotesize{Isosurface maps of the three observables: size $z_1$, primary distortion $z_2$, and secondary distortion $z_3$. The panels show how these observables depend on the parameters $(a,i,r_{nl})$. In each panel, differently colored surfaces indicate isosurfaces corresponding to different values of the observable. This parameter dependence forms the basis for determining $(a,i,r_{nl})$ from the observables $(z_1,z_2,z_3)$.}}
	\label{fig050}
\end{figure}
\begin{figure}[tb]
	\begin{tabular}{ ccc }
		\includegraphics[height=4.3cm]{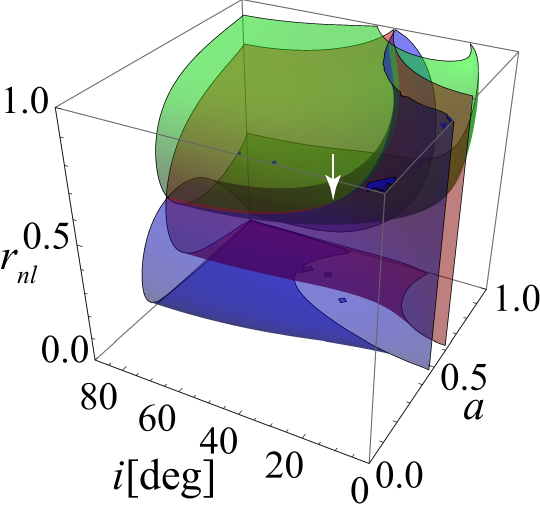} &
		\includegraphics[height=4.3cm]{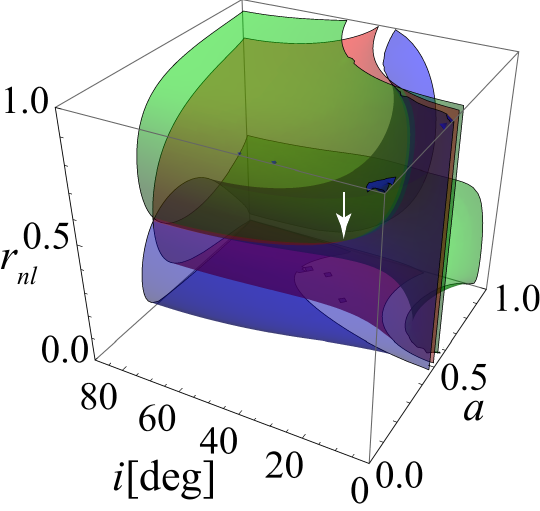} &
		\includegraphics[height=4.7cm]{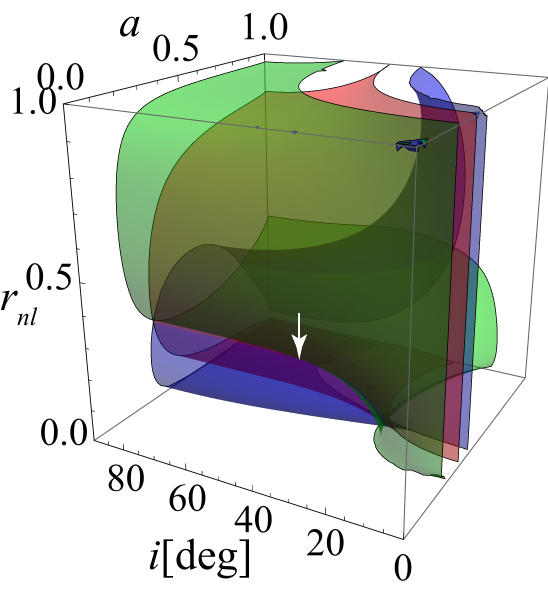} \\
		(a) $z_2=0.05913$, $z_3=0.04423$ & (b) $z_2=0.05915$, $z_3=0.04423$ & (c) $z_2=0.05917$, $z_3=0.04423$ \\
		\includegraphics[height=4.3cm]{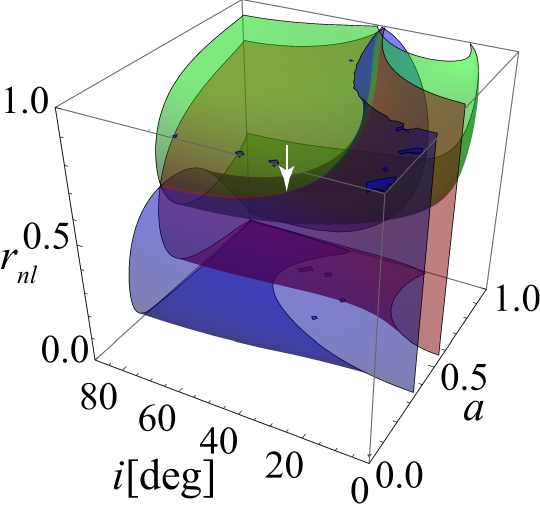} &
		\includegraphics[height=4.3cm]{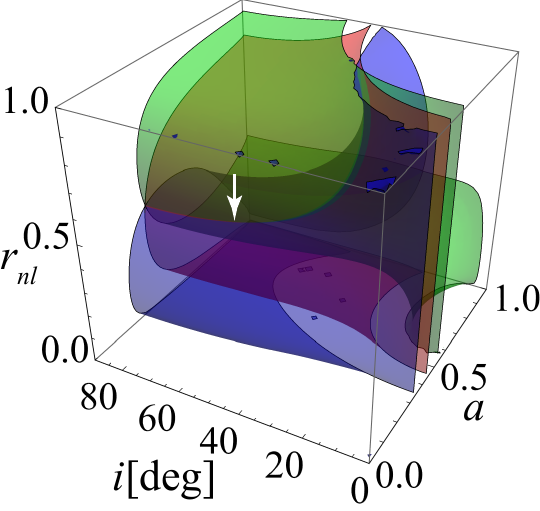} &
		\includegraphics[height=4.7cm]{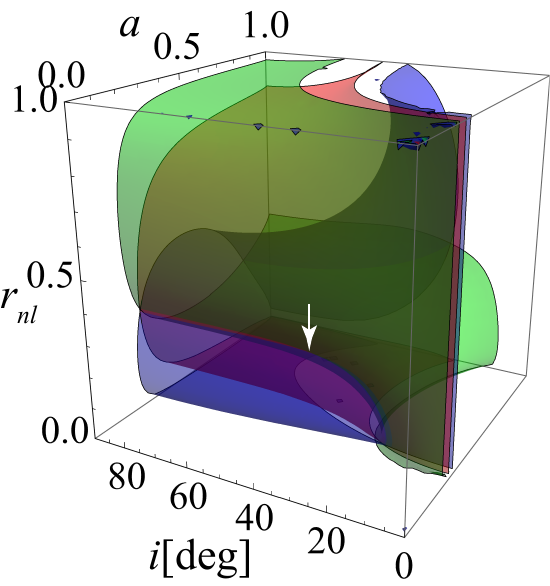} \\
		(d) $z_2=0.05913$, $z_3=0.044215$ & (e) $z_2=0.05915$, $z_3=0.044215$ & (f) $z_2=0.05917$, $z_3=0.044215$ \\
		\includegraphics[height=4.3cm]{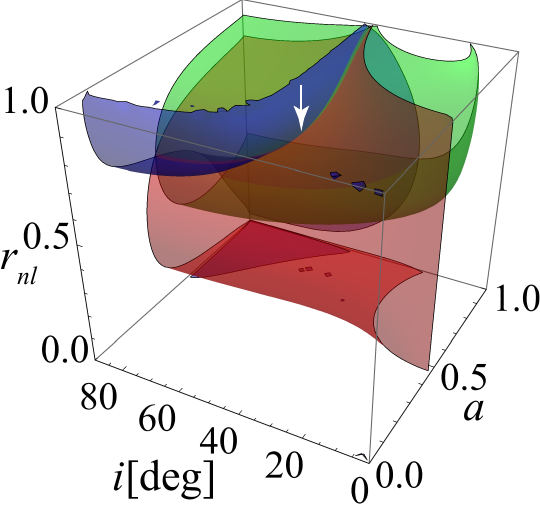} &
		\includegraphics[height=4.3cm]{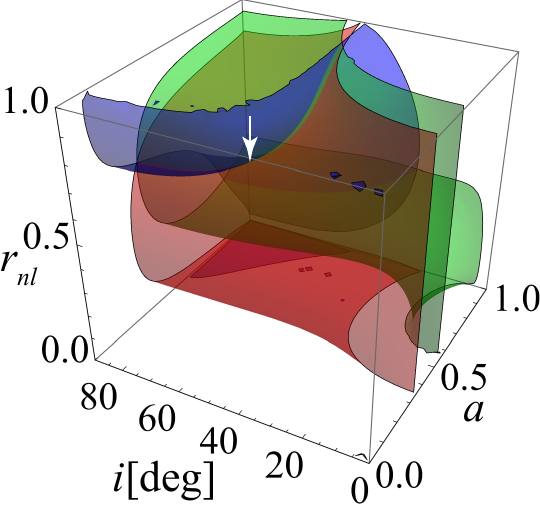} &
		\includegraphics[height=4.3cm]{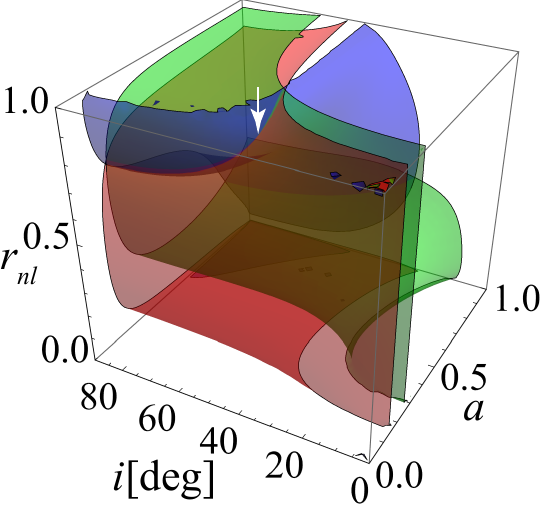}  \\
		(g) $z_2=0.05913$, $z_3=0.044195$ & (h) $z_2=0.05915$, $z_3=0.044195$ & (i) $z_2=0.05917$, $z_3=0.044195$ \\
	\end{tabular}
	\caption{\footnotesize{Combined isosurface maps of the three observables: size $z_1$ (red), primary distortion $z_2$ (green), and secondary distortion $z_3$ (blue). The nine panels (a)--(i) show how the three isosurfaces change as the observable values are varied. Within each row, $z_3$ is fixed while the isosurface of $z_2$ is varied. Within each column, $z_2$ is fixed while the isosurface of $z_3$ is varied. In panels (c) and (f), the parameter-space box is viewed from a lower angle than in the other panels.
This is because, from a higher viewpoint, the intersection point of the three isosurfaces is obscured by the bulge of the green isosurface. In every panel, the three isosurfaces intersect at a single point, indicated by the white arrow. This suggests that the mapping from the parameters $(a,i,r_{nl})$ to the observables $(z_1,z_2,z_3)$ is one-to-one. Panels (a)--(i) correspond to $z_1 = 0.1411$, $0.14025$, $0.13955$, $0.1406$, $0.13985$, $0.13917$, $0.13995$, $0.139303$, and $0.13869$, respectively.}}
	\label{fig090}
\end{figure}
We propose the use of observables to distinguish between different
standardized segments of the critical curve.
An observable is defined as a measurable quantity derived from a standardized segment of the critical curve that captures its essential geometric characteristics.
In particular, we define three observables characterizing the standardized segment of the critical curve.
These observables are constructed from the Fourier coefficients of the
standardized segment of the critical curve, as described below.

As shown in Fig.~\ref{fig020}, we rotate and translate the screen
coordinates so that the two endpoints of a standardized segment of the critical curve are fixed at
$(-1,0)$ and $(1,0)$.
We then introduce polar coordinates and represent the standardized segment of the critical curve by a
radial function
\begin{eqnarray}
\rho(\varphi; a , i , r_{nl}).
\end{eqnarray}
The function $\rho(\varphi; a , i , r_{nl})$ is defined on
$0 \le \varphi \le \pi$ and extended to $-\pi \le \varphi \le \pi$
by imposing the symmetry condition
\begin{eqnarray}
\rho(-\varphi)=\rho(\varphi).
\end{eqnarray}

To systematically construct observables, we consider the complex
Fourier-series representation
\begin{eqnarray}
\rho(\varphi; a , i , r_{nl})
&=&
\sum_{n=-\infty}^{\infty}
c_n(a,i,r_{nl})\,e^{{\rm i}n\varphi},
\\
c_n(a,i,r_{nl})
&=&
\frac{1}{2\pi}
\int_{-\pi}^{\pi}
\rho(\varphi; a,i,r_{nl})
e^{-{\rm i}n\varphi}\,{\rm d}\varphi .
\end{eqnarray}
Since $\rho(\varphi)$ is a real-valued even function,
the Fourier coefficients satisfy $c_{-n}=c_n$ and are real.
Moreover, the piecewise smoothness of $\rho(\varphi)$~\cite{Paganini:2017qfo} ensures convergence of the Fourier series.

We define the truncated Fourier-coefficient vector by
\begin{eqnarray}
\boldsymbol{c}(a,i,r_{nl})
=
\begin{pmatrix}
c_0 \\ c_1 \\ \vdots \\ c_7
\end{pmatrix}.
\end{eqnarray}

%-------------------------------------------%
\subsection{Principal-component observables}
%-------------------------------------------%
Suppose that $(c_0,c_1,c_2)$ are chosen as three observables, and
consider the mapping
\begin{eqnarray}
(a,i,r_{nl}) \mapsto (c_0,c_1,c_2).
\end{eqnarray}
Although we do not discuss this point in detail here, our numerical analysis does not establish the injectivity of this mapping. We therefore incorporate information from the Fourier coefficients of degree $3$ and higher into the observables. Accordingly, we define three quantities as linear combinations of the Fourier coefficients up to seventh order.
Specifically, we apply an orthogonal transformation to the vector $\boldsymbol{c}$
and take the first three components of the transformed vector as
observables characterizing the standardized segment of the critical curve.
This transformation is determined using principal component analysis
(PCA)~\cite{Jolliffe}, which constructs orthogonal variables
(principal components) that successively maximize the variance
of a multivariate data set.

We uniformly sample $50^3$ points from the parameter space
$P \times [0,1]$ and construct a data set $F$
consisting of the corresponding Fourier-coefficient vectors.
Applying PCA to $F$, we obtain an orthogonal transformation
that maximizes the variance of the sampled data.
The resulting orthogonal matrix ${\bm A}$ is provided as
Supplemental Material~\cite{SupplementalMaterial1} to ensure reproducibility.

We define the principal-component vector $\boldsymbol{z}$ by
\begin{eqnarray}
\boldsymbol{z}(a,i,r_{nl})
=
\begin{pmatrix}
z_1 \\ z_2 \\ \vdots \\ z_8
\end{pmatrix}
=
{\bm A}^{\mathsf T}\boldsymbol{c}.
\end{eqnarray}
Applying this transformation to the data set $F$, we obtain the
corresponding data set $Z$ of principal components.
For reproducibility, we also provide the combined data set
$(F,Z)$ indexed by $(a,i,r_{nl})$ as Supplemental Material~\cite{SupplementalMaterial2}.

The isosurfaces of the observables $z_1$, $z_2$, and $z_3$
are shown in Fig.~\ref{fig050}.

%-------------------------------------------%
\section{Parameter Determination Based on Observables}
\label{sec:pm}
%-------------------------------------------%
Figure~\ref{fig050}(a) shows that, near $a=0.5$, the portion of the
red isosurface facing the $(i,r_{nl})$ plane becomes nearly parallel to that plane, and Fig.~\ref{fig050}(b) shows that the same tendency appears near $a=0.3$.
These behaviors can be understood from the Schwarzschild limit.
When $a=0$, the critical curve reduces to a circle, and all standardized segments of the critical curve become identical arcs.
Consequently, the standardized segments of the critical curve are independent of both the inclination angle $i$ and the parameter $r_{nl}$.
Therefore, their Fourier coefficients take the same values for all $(i,r_{nl})$, and hence the observables $z_1$, $z_2$, and $z_3$, which are defined as linear combinations of these coefficients, are also independent of $i$ and $r_{nl}$.
Accordingly, at $a=0$ the isosurfaces of these observables are parallel to the $(i,r_{nl})$ plane.
Figure~\ref{fig050}(a) suggests that this tendency already appears around $a=0.5$ for $z_1$, while Fig.~\ref{fig050}(b) suggests that it already appears around $a=0.3$ for $z_2$.
By contrast, Fig.~\ref{fig050}(c) does not yet show a comparable tendency around $a=0.3$ for $z_3$.

As can be seen from the transformation matrix $\bm{A}$ provided as Supplemental Material~\cite{SupplementalMaterial1}, although $z_1$ is a linear combination of eight Fourier coefficients, the coefficient of $c_0$ is about ten times larger than those of the other coefficients. Furthermore, the magnitude of $c_0$ is larger than those of the other seven Fourier coefficients. Therefore, $z_1$ is dominated by $c_0$. Since $c_0$ is the average of $\rho(\varphi)$, we refer to $z_1$ as the size.

As shown in Fig.~\ref{fig050}(b), in the range $0.3 \le a \le 0.6$, the portions of the red and blue isosurfaces facing the $(i,r_{nl})$ plane extend mainly along the $i$ and $r_{nl}$ directions. This suggests that, in the range $0.3 \le a \le 0.6$, the value of $z_2$ depends only weakly on $i$ and $r_{nl}$ and is therefore governed primarily by $a$. We therefore interpret $z_2$ as characterizing the dependence of the segment shape on $a$. We refer to $z_2$ as the primary distortion.

Figure~\ref{fig050}(c) shows that the purple isosurface extends mainly along the $r_{nl}$ direction. This indicates that, near this isosurface, $z_3$ depends only weakly on $r_{nl}$ and depends on both $a$ and $i$. We regard this tendency as a characteristic feature of $z_3$. We refer to $z_3$ as the secondary distortion.

These geometric properties of the observables provide the basis for parameter determination.

Varying the values of the three observables $(z_1, z_2, z_3)$,
we construct the corresponding isosurfaces in Fig.~\ref{fig090}.
As shown in Fig.~\ref{fig090}, in all sampled cases the three
isosurfaces intersect at a single point.
This strongly suggests that the map
\begin{eqnarray}
(a,i,r_{nl}) \mapsto (z_1,z_2,z_3)
\end{eqnarray}
is injective over the considered domain.
In other words, once a standardized segment of the critical curve is obtained observationally,
the parameters $(a, i, r_{nl})$
can be uniquely determined within this framework.
This injectivity indicates the absence of parameter degeneracy
in the considered parameter region.

Consequently, if the black hole mass $M$ is known a priori from independent
observations, all three physical parameters $(M, aM, i)$ can, in principle,
be determined.

%-------------------------------------------%
\section{Accuracy of the parameter-determination method}
\label{sec:demo}
%-------------------------------------------%
%-------------------------------------------%
\subsection{Accuracy and error distribution}
\label{sec:numerical}
%-------------------------------------------%
\begin{table}[tb]
\caption{\footnotesize{
Errors of the parameter-determination method evaluated without
artificial perturbations in the observables.  The RMSE and MAE are shown
with their $99\%$ bootstrap confidence intervals in brackets.
}}
\label{tab:reconstruction_errors_noiseless}
\begin{ruledtabular}
\begin{tabular}{ccc}
Parameter
&
RMSE
&
MAE
\\
\hline
$a$
&
\begin{tabular}{@{}c@{}}
$2.4059\times10^{-3}$ \\
$[7.7665\times10^{-4},\,3.9206\times10^{-3}]$
\end{tabular}
&
\begin{tabular}{@{}c@{}}
$2.5954\times10^{-4}$ \\
$[1.0814\times10^{-4},\,4.8774\times10^{-4}]$
\end{tabular}
\\
$i$ [deg]
&
\begin{tabular}{@{}c@{}}
$1.2511$ \\
$[4.6326\times10^{-1},\,1.9655]$
\end{tabular}
&
\begin{tabular}{@{}c@{}}
$1.3025\times10^{-1}$ \\
$[4.7472\times10^{-2},\,2.4668\times10^{-1}]$
\end{tabular}
\\
$r_{nl}$
&
\begin{tabular}{@{}c@{}}
$9.1210\times10^{-3}$ \\
$[3.9475\times10^{-3},\,1.3924\times10^{-2}]$
\end{tabular}
&
\begin{tabular}{@{}c@{}}
$1.0400\times10^{-3}$ \\
$[4.1966\times10^{-4},\,1.8663\times10^{-3}]$
\end{tabular}
\\
\end{tabular}
\end{ruledtabular}
\end{table}
\begin{figure}[tb]
	\begin{tabular}{ccc}
	\includegraphics[height=4.5cm]{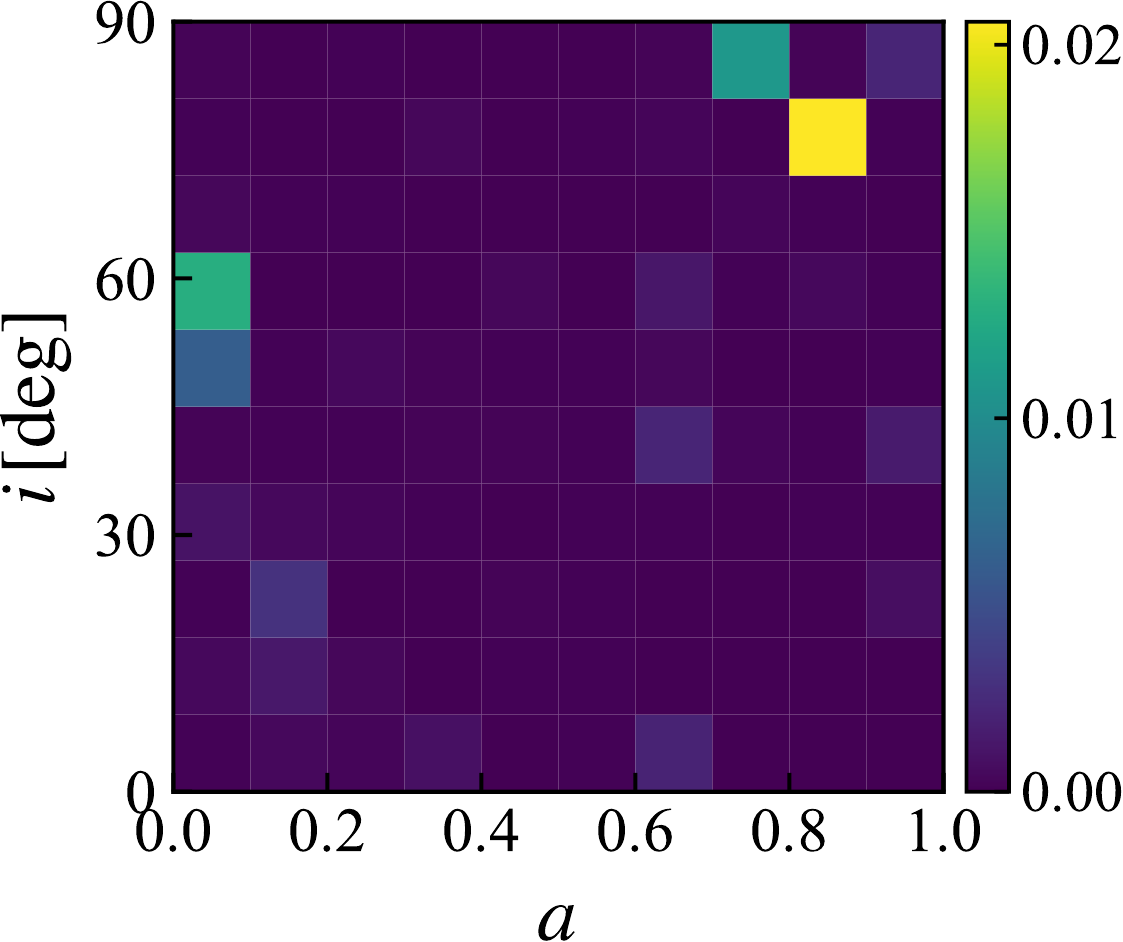} &
	\includegraphics[height=4.5cm]{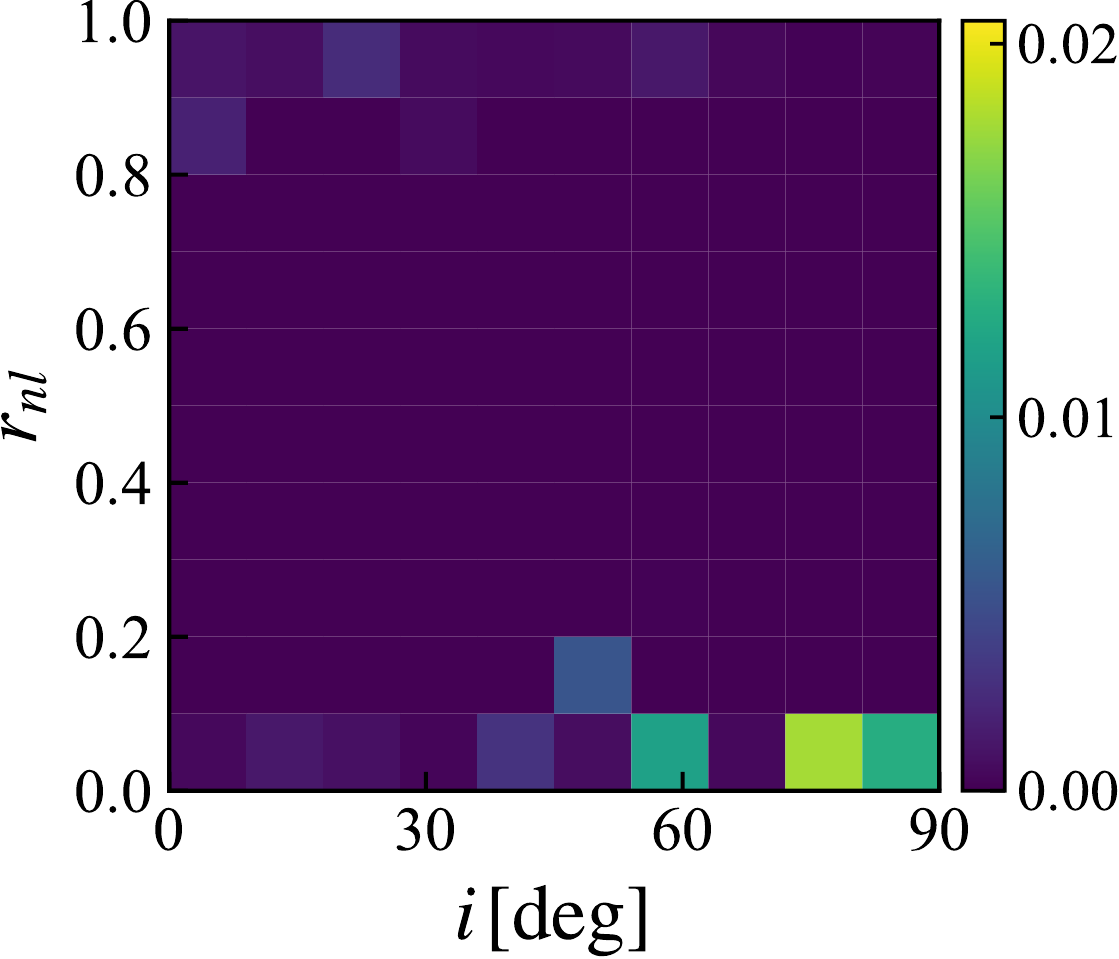} &
	\includegraphics[height=4.5cm]{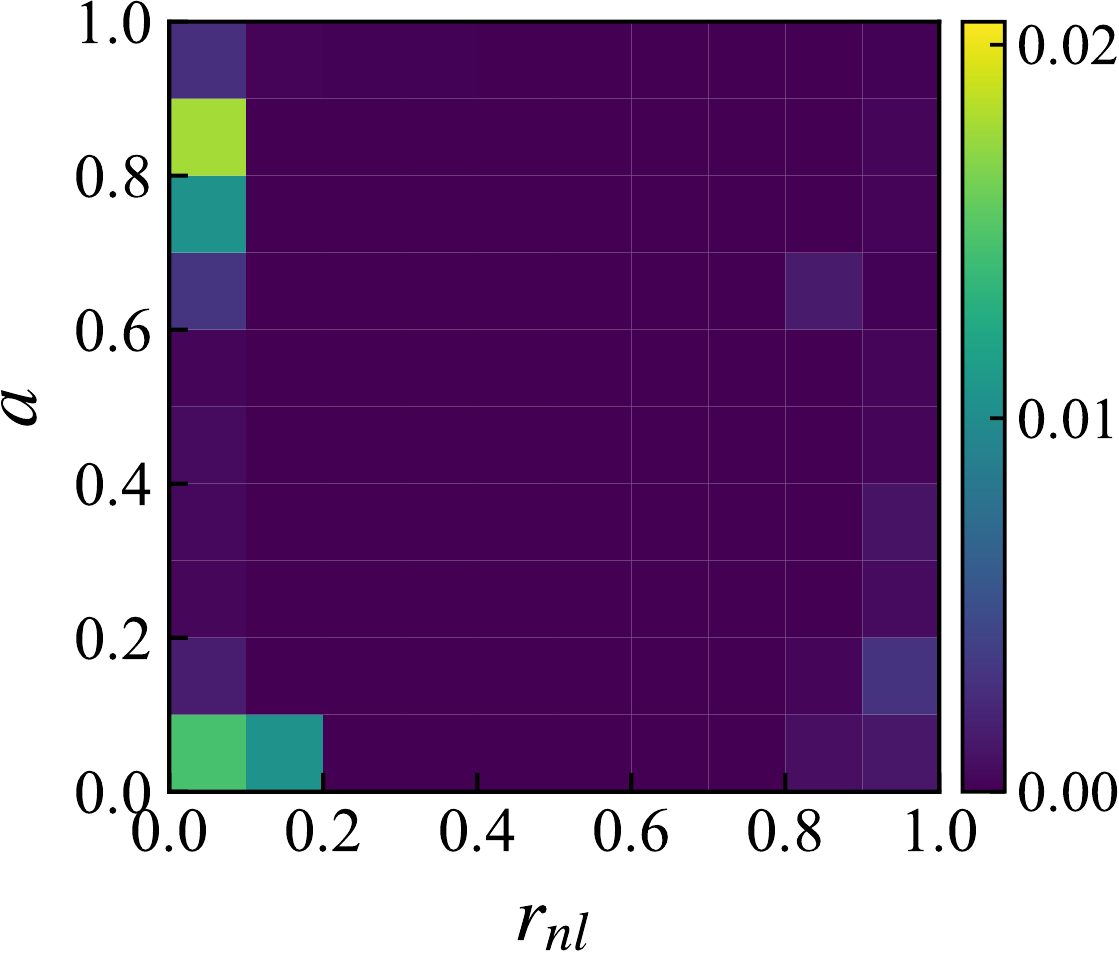} \\
	(a) $a$ error, $(a,i)$ &
	(b) $a$ error, $(i,r_{nl})$ &
	(c) $a$ error, $(r_{nl},a)$ \\
	\includegraphics[height=4.5cm]{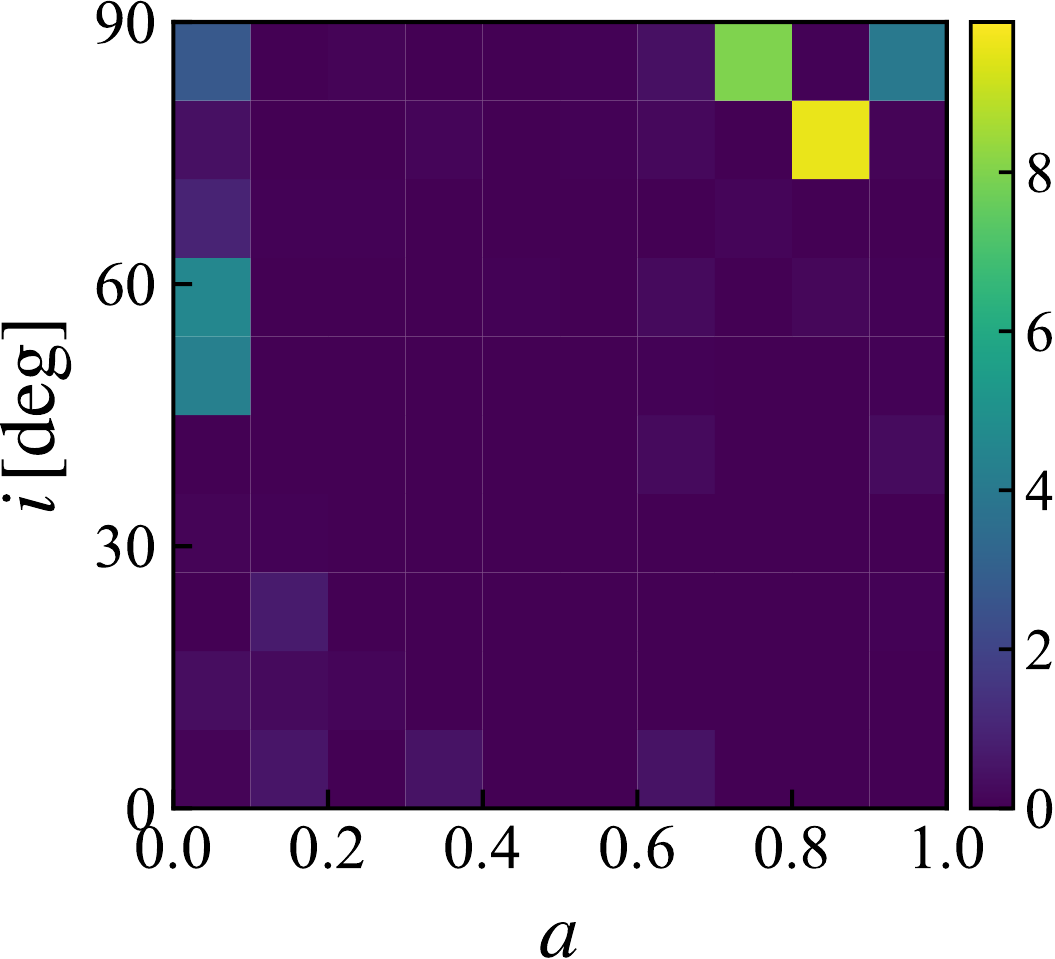} &
	\includegraphics[height=4.5cm]{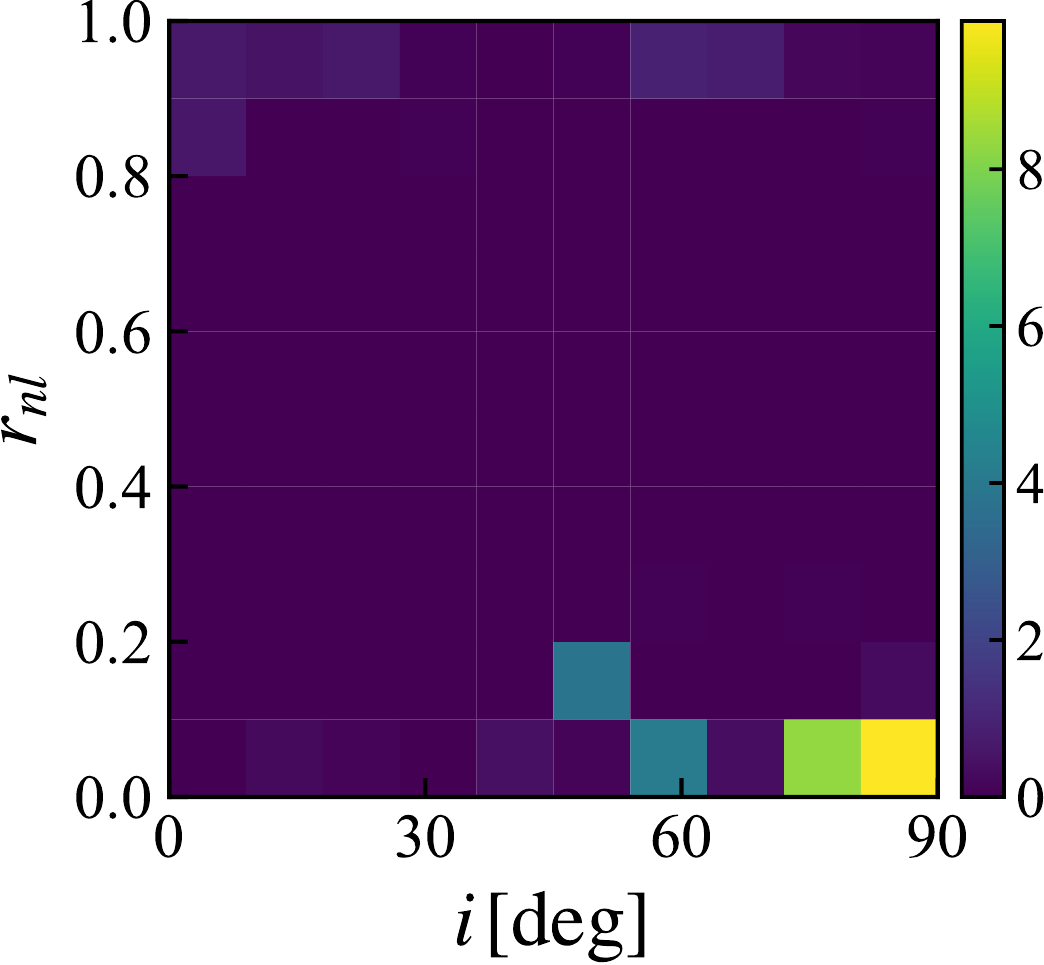} &
	\includegraphics[height=4.5cm]{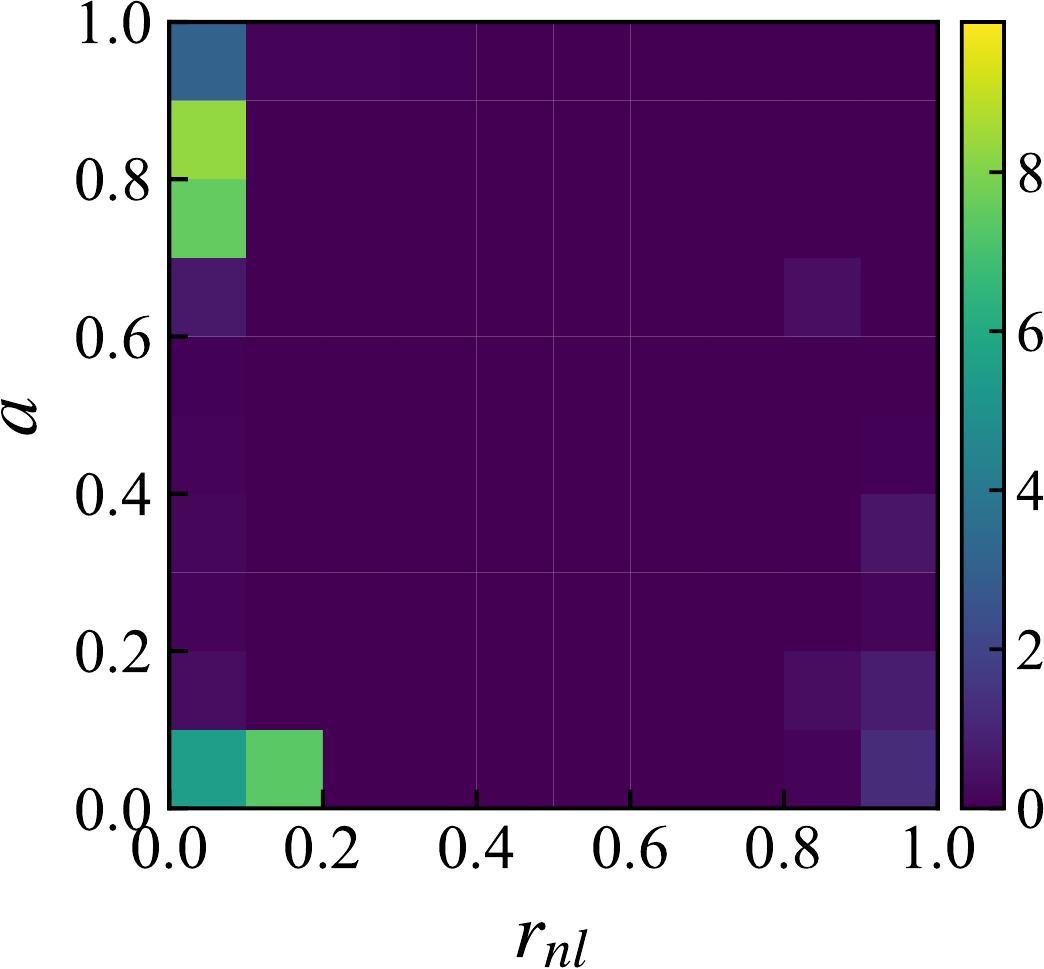} \\
	(d) $i$ error, $(a,i)$ &
	(e) $i$ error, $(i,r_{nl})$ &
	(f) $i$ error, $(r_{nl},a)$ \\
	\includegraphics[height=4.5cm]{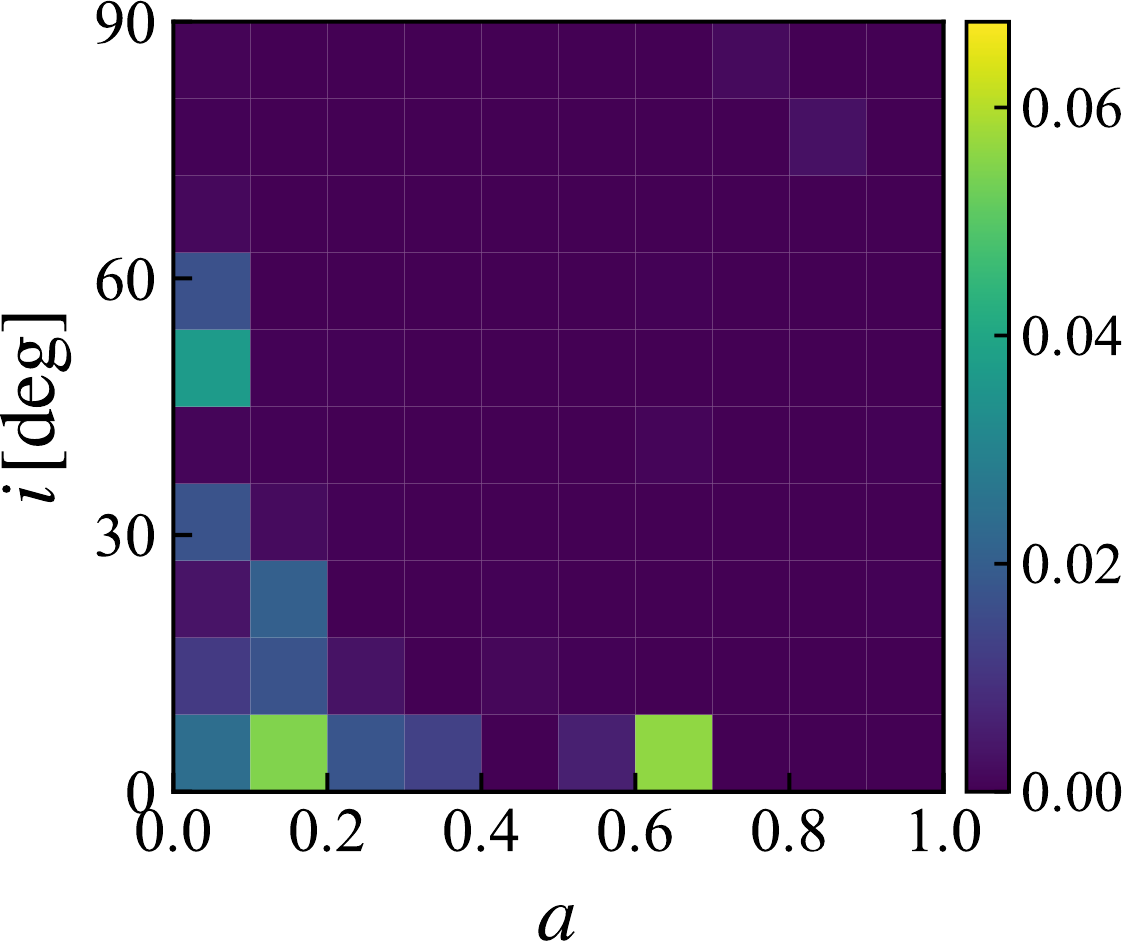} &
	\includegraphics[height=4.5cm]{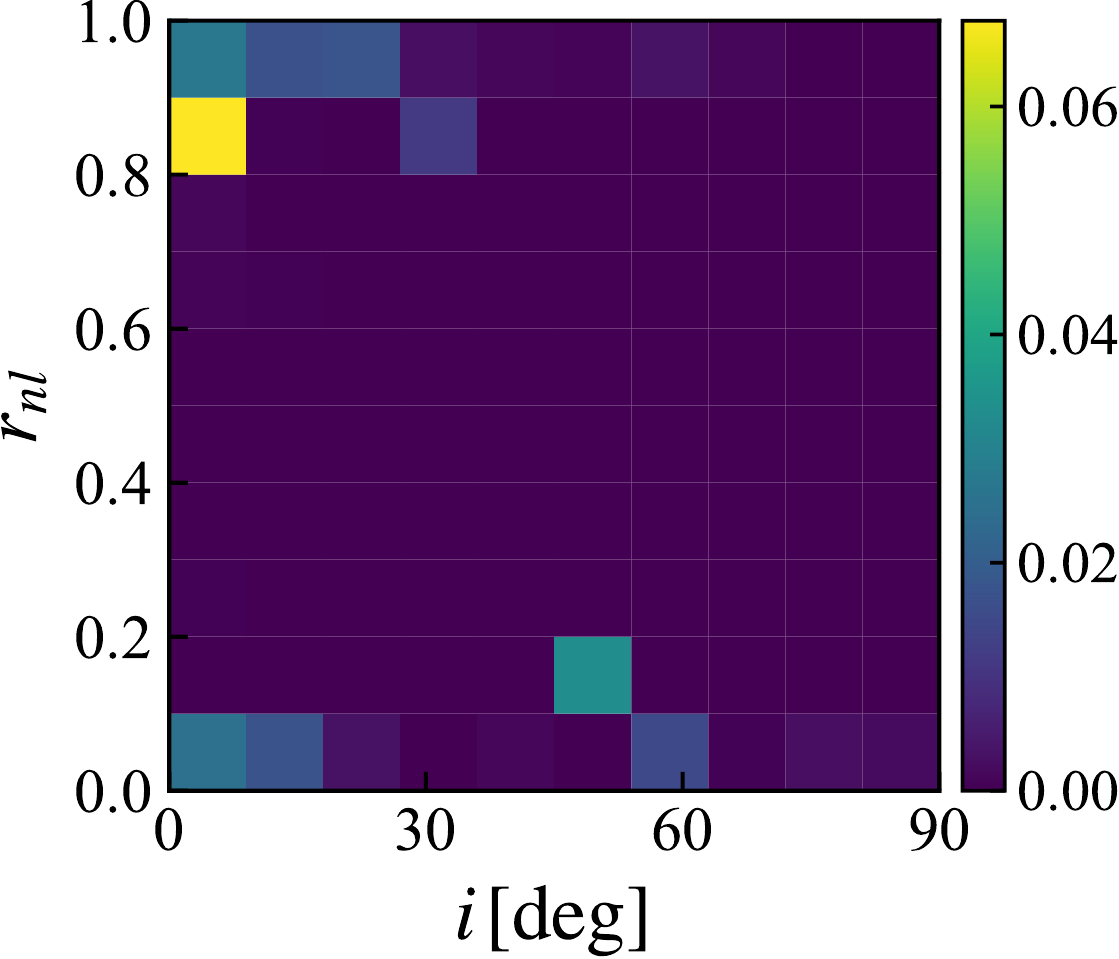} &
	\includegraphics[height=4.5cm]{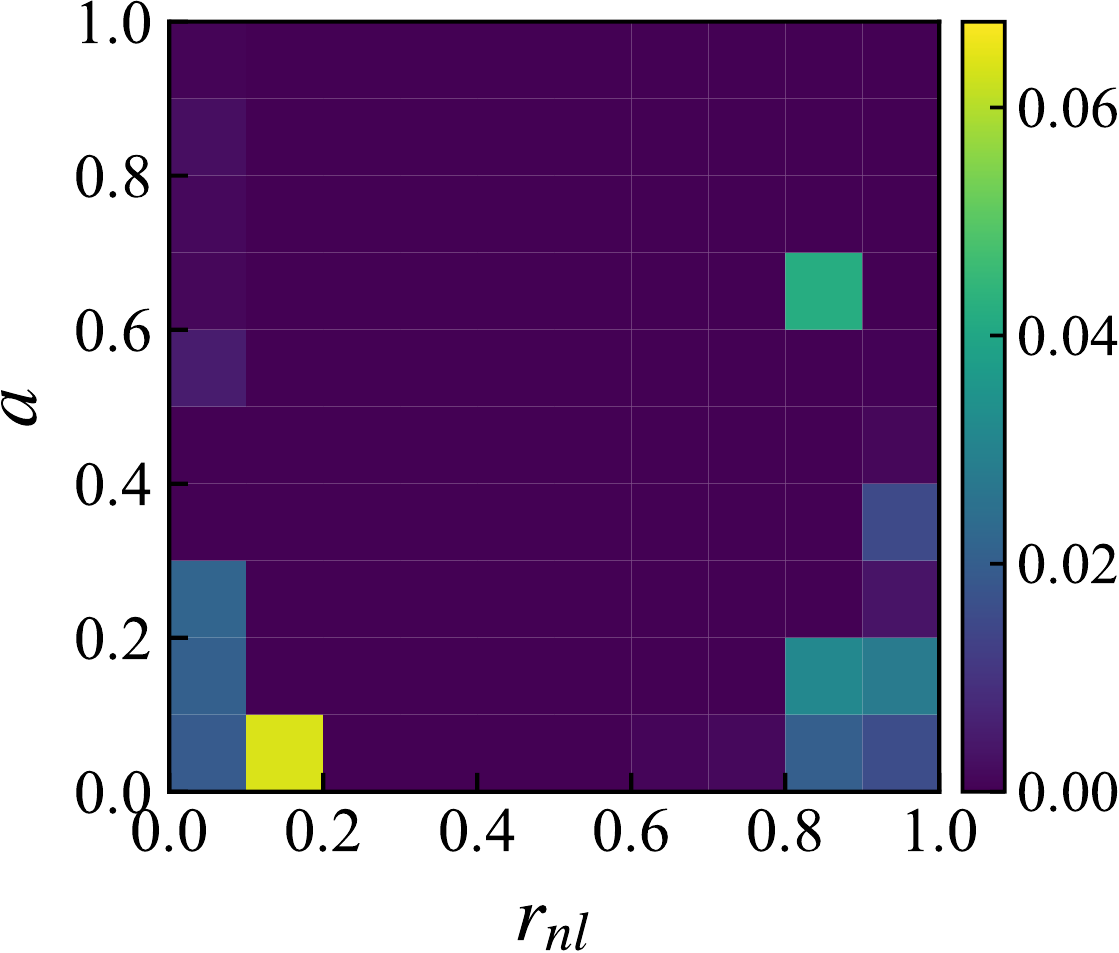} \\
	(g) $r_{nl}$ error, $(a,i)$ &
	(h) $r_{nl}$ error, $(i,r_{nl})$ &
	(i) $r_{nl}$ error, $(r_{nl},a)$ \\
	\end{tabular}
	\caption{\footnotesize{Local RMSE maps for the reconstructed parameters $a$, $i$, and $r_{nl}$.
Panels (a)--(c), (d)--(f), and (g)--(i) show the local RMSE of
$a$, $i$, and $r_{nl}$, respectively, in the parameter planes
$(a,i)$, $(i,r_{nl})$, and $(r_{nl},a)$.  In each parameter plane, the
evaluation samples are divided into $10^2$ bins, and the local RMSE is
computed for each bin.  The color scale represents the local RMSE of
the parameter shown in each row.}}
	\label{fig100}
\end{figure}
In this subsection, we evaluate the numerical accuracy of the proposed
parameter-determination method under idealized conditions, without
including observational uncertainties associated with the identification
of segments.  The stability of the method against perturbations in the
observables is examined in Sec.~\ref{sec:sta}.

First, we sample parameter values $(a,i,r_{nl})$ from the parameter space and generate a segment corresponding to these values. We then compute the Fourier coefficient vector $\boldsymbol{c}$ of this segment. Using the transformation matrix $\bm A$ provided in Supplemental Material~\cite{SupplementalMaterial1}, we compute the principal-component vector $\boldsymbol{z}$ as $\boldsymbol{z}=\bm{A}^{\mathsf T}\boldsymbol{c}$. The first three components of $\boldsymbol{z}$ correspond to the observables $(z_1,z_2,z_3)$.

The parameter values are then reconstructed from the observables $(z_1,z_2,z_3)$. Using the dataset provided as Supplemental Material~\cite{SupplementalMaterial2}, we construct the isosurfaces of the observables in the $(a,i,r_{nl})$ parameter space. The intersection point of these isosurfaces determines the corresponding parameter values.

Numerically, this intersection point is obtained by implementing the inverse mapping from $(z_1,z_2,z_3)$ to $(a,i,r_{nl})$ based on Supplemental Material~\cite{SupplementalMaterial2}. The data are first arranged on a three-dimensional grid in the $(a,i,r_{nl})$ parameter space, and interpolating functions for $(z_1,z_2,z_3)$ are constructed. By finding the interpolated observable values that minimize the squared
difference from the given observable values, we obtain the corresponding parameter values as those associated with the given observable values.\footnote{The numerical implementation was carried out in Python using SciPy; in particular, grid interpolation and nonlinear least-squares optimization were employed.} As shown in Sec.~\ref{sec:pm}, the mapping from $(a,i,r_{nl})$ to $(z_1,z_2,z_3)$ is injective, and therefore the corresponding parameter values are uniquely determined by a given set of observables $(z_1,z_2,z_3)$. In the reconstruction itself, the true sampled parameter values are not used; they serve only as a reference for evaluation after the reconstruction.

The reconstructed parameter values are compared with the corresponding true sampled parameter values, and the resulting errors are computed. This process is then repeated for $10^3$ sampled points in parameter space to assess the accuracy of the parameter-determination method.

For this evaluation, we first construct a $40^3$ grid in parameter space and then randomly select $10^3$ samples from it. Since this evaluation grid is distinct from the $50^3$ grid used in Supplemental Material~\cite{SupplementalMaterial2}, the resulting errors reflect interpolation performance rather than trivial reconstruction at the tabulated points.

As error indicators, we use the root mean squared error (RMSE) and the mean absolute error (MAE). The RMSE is defined as the square root of the mean of the squared differences between the true and reconstructed parameter values, whereas the MAE is defined as the mean of the absolute differences between them. Smaller values of RMSE and MAE indicate higher reconstruction accuracy. Compared with the RMSE, the MAE is less sensitive to samples for which the differences between the reconstructed and true parameter values are large.

Using $10^3$ evaluation samples, we quantified the reconstruction accuracy by computing the RMSE and MAE as shown in Table~\ref{tab:reconstruction_errors_noiseless}. The method achieved RMSE values of $2.4059\times10^{-3}$, $1.2511~[\mathrm{deg}]$, and $9.1210\times10^{-3}$ for $a$, $i$, and $r_{nl}$, respectively, with corresponding $99\%$ bootstrap confidence intervals of $[7.7665\times10^{-4},\,3.9206\times10^{-3}]$, $[4.6326\times10^{-1},\,1.9655]~[\mathrm{deg}]$, and $[3.9475\times10^{-3},\,1.3924\times10^{-2}]$. The corresponding MAE values were $2.5954\times10^{-4}$, $1.3025\times10^{-1}~[\mathrm{deg}]$, and $1.0400\times10^{-3}$, with $99\%$ bootstrap confidence intervals of $[1.0814\times10^{-4},\,4.8774\times10^{-4}]$, $[4.7472\times10^{-2},\,2.4668\times10^{-1}]~[\mathrm{deg}]$, and $[4.1966\times10^{-4},\,1.8663\times10^{-3}]$, respectively. Here, the bootstrap confidence interval is defined by the 0.5th and 99.5th percentiles of the bootstrap distribution obtained by repeatedly resampling the $10^3$ evaluation samples with replacement and recalculating the RMSE or MAE. This distribution quantifies the uncertainty in the estimated RMSE and MAE due to the limited number of evaluation samples.

Since the RMSE weights larger errors more strongly than the MAE, the
difference between the RMSE and MAE provides information on the spread
of the error distribution.  For the values listed in
Table~\ref{tab:reconstruction_errors_noiseless}, this difference is
particularly pronounced for the inclination angle $i$, suggesting that
the error distribution for $i$ includes a limited number of samples with
relatively large errors.

To examine the parameter dependence of the errors, we show in
Fig.~\ref{fig100} the local RMSE maps for the reconstructed parameters
$a$, $i$, and $r_{nl}$ in the $(a,i)$, $(i,r_{nl})$, and $(r_{nl},a)$
planes.  Each plane is divided into $10^2$ bins, and the local RMSE of
each parameter is computed in each bin from the differences between the
reconstructed and true values for the samples contained in that bin.
The local RMSE therefore measures the typical local error of $a$, $i$,
or $r_{nl}$ in each region of the parameter plane.

We use the 99th percentiles of the absolute errors as reference scales
for interpreting the local RMSE maps.  These values are
$4.1279\times10^{-3}$, $1.7033~[\mathrm{deg}]$, and
$4.2082\times10^{-2}$ for $a$, $i$, and $r_{nl}$, respectively.
We regard the local RMSE of each parameter as relatively enhanced when it
exceeds the corresponding 99th-percentile reference scale. The local RMSE of the spin parameter $a$ is relatively enhanced only in localized regions, mainly
in the high-inclination and small-$r_{nl}$ region.
For the inclination angle $i$, the local RMSE shows more pronounced relative enhancements
in the same region.  By contrast, the local RMSE of $r_{nl}$ is
relatively enhanced mainly in the low-spin and low-inclination region.

Relatively enhanced local RMSE values for $a$ and $i$ are found in the
high-inclination and small-$r_{nl}$ region.  The error enhancement in
this part of the parameter space can be understood from the geometry of
the critical curve.  When $r_{nl}$ is small, the segment is located near
the left end of the critical curve, as shown in Fig.~\ref{fig030}.  As
the inclination angle $i$ increases, the left part of the critical curve
becomes more distorted and develops a concave shape~\cite{Hioki:2009na}.
In this parameter region, the standardized segment approaches a nearly straight
segment.  As a result, different parameter values can give rise to
similar standardized segments and hence similar observable values,
making the reconstruction less accurate there.

The relative enhancement of the local RMSE of $r_{nl}$ in the low-spin
and low-inclination region can be understood from the fact that the
critical curve becomes closer to a circle in this region~\cite{Hioki:2009na}.  As a result,
changing the position of the segment along the critical curve produces
only a weak change in the standardized segment.  Different values of
$r_{nl}$ can therefore give rise to similar observable values, making the reconstruction of $r_{nl}$ less accurate.

The pronounced local RMSE enhancement for $i$ is qualitatively
consistent with the isosurface structure shown in Fig.~\ref{fig090}.
There are three pairwise intersections among the three isosurfaces.  In
some regions, these intersection curves extend in the direction of the
inclination angle $i$.  Although the three isosurfaces intersect at a
single point, the corresponding pairwise intersection curves run nearly
parallel to each other with small separations in the $i$ direction.
This structure indicates that the constraints on $i$ are relatively weak
in such regions.  As a result, small changes in the observable values
can lead to relatively large changes in the reconstructed value of $i$.

To assess the RMSE and local RMSE values in Fig.~\ref{fig100}, we
introduce two reference scales.  The first is an uninformative baseline,
defined by assuming that the true parameter value is uniformly
distributed over its allowed range, while the reconstructed value is
always fixed to the central value of that range.  For a parameter whose
allowed range has length $L$, the RMSE of this baseline is
$L/\sqrt{12}$, because it is equal to the standard deviation of a
uniform distribution over that range.  Thus, for the parameters $a$ and
$r_{nl}$, whose ranges have length unity, the uninformative baseline is
$1/\sqrt{12}\simeq0.28868$.  For the inclination angle $i$, which is
measured in degrees and lies in the range $0<i\leq90$, the analogous
baseline corresponds to always choosing $45~[\mathrm{deg}]$, giving an
RMSE of $90/\sqrt{12}\simeq25.981~[\mathrm{deg}]$.
Since there is no absolute threshold for judging whether an error is sufficiently small, in this paper we define the informative baseline of each parameter as
one-tenth of the corresponding uninformative baseline.

The RMSE values for $a$, $i$, and $r_{nl}$ shown in Table~\ref{tab:reconstruction_errors_noiseless} are below the informative baselines.
Moreover, the local RMSE of $a$ remains below the informative baseline
throughout the parameter space.
The local RMSE values of $i$ and $r_{nl}$ are comparable with the corresponding uninformative baselines in localized regions: the high-inclination and small-$r_{nl}$ region for $i$, and
the low-spin and low-inclination region for $r_{nl}$.

%-------------------------------------------%
\subsection{Stability against perturbations in the observables}
\label{sec:sta}
%-------------------------------------------%
\begin{table}[tb]
\caption{\footnotesize{
Errors of the parameter-determination method after artificial Gaussian
perturbations are added to the observables $(z_1,z_2,z_3)$.  The
perturbation amplitude $\sigma$ is measured relative to the standard
deviation of each observable over the $10^3$ evaluation samples.
}}
\label{tab:reconstruction_errors_perturbed}
\begin{ruledtabular}
\begin{tabular}{cccc}
	$\sigma$ & Parameter & RMSE & MAE \\
	\hline
	$10^{-1}$ & $a$      & $1.7286\times10^{-1}$ & $1.2618\times10^{-1}$ \\
           & $i$ [deg] & $36.6409$              & $28.6574$              \\
           & $r_{nl}$ & $3.1303\times10^{-1}$ & $2.1917\times10^{-1}$ \\
	\hline
	$10^{-3}$ & $a$      & $1.4043\times10^{-2}$ & $8.7051\times10^{-3}$ \\
           & $i$ [deg] & $13.3180$              & $6.0592$               \\
           & $r_{nl}$ & $1.0033\times10^{-1}$ & $4.2859\times10^{-2}$ \\
	\hline
	$10^{-4}$ & $a$      & $5.1171\times10^{-3}$ & $2.4598\times10^{-3}$ \\
           & $i$ [deg] & $6.7823$               & $2.2081$               \\
           & $r_{nl}$ & $5.7628\times10^{-2}$ & $1.8369\times10^{-2}$ \\
	\hline
	$10^{-8}$ & $a$      & $4.5133\times10^{-3}$ & $1.5831\times10^{-3}$ \\
           & $i$ [deg] & $4.4822$               & $1.4070$               \\
           & $r_{nl}$ & $4.5174\times10^{-2}$ & $1.2532\times10^{-2}$ \\
\end{tabular}
\end{ruledtabular}
\end{table}
\begin{figure}[tb]
	\begin{tabular}{ccc}
	\includegraphics[height=4.5cm]{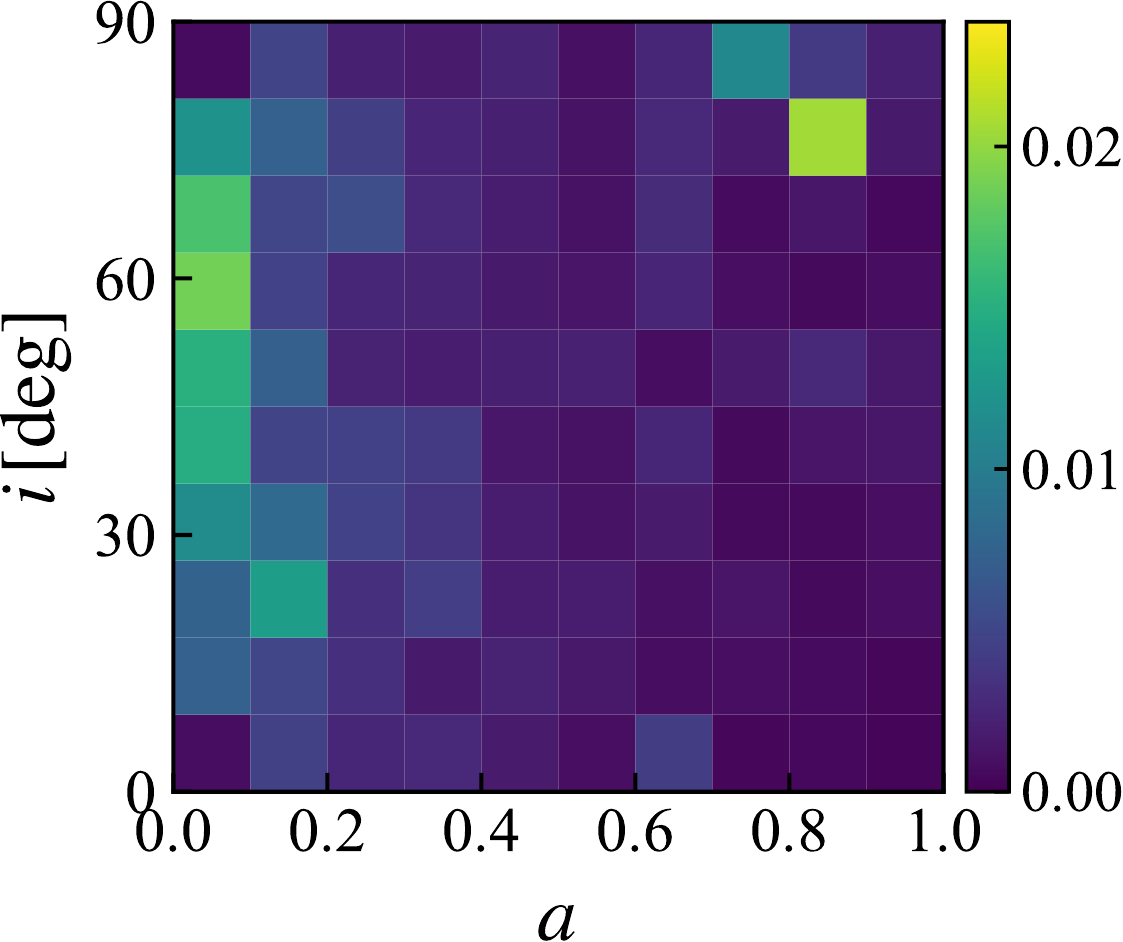} &
	\includegraphics[height=4.5cm]{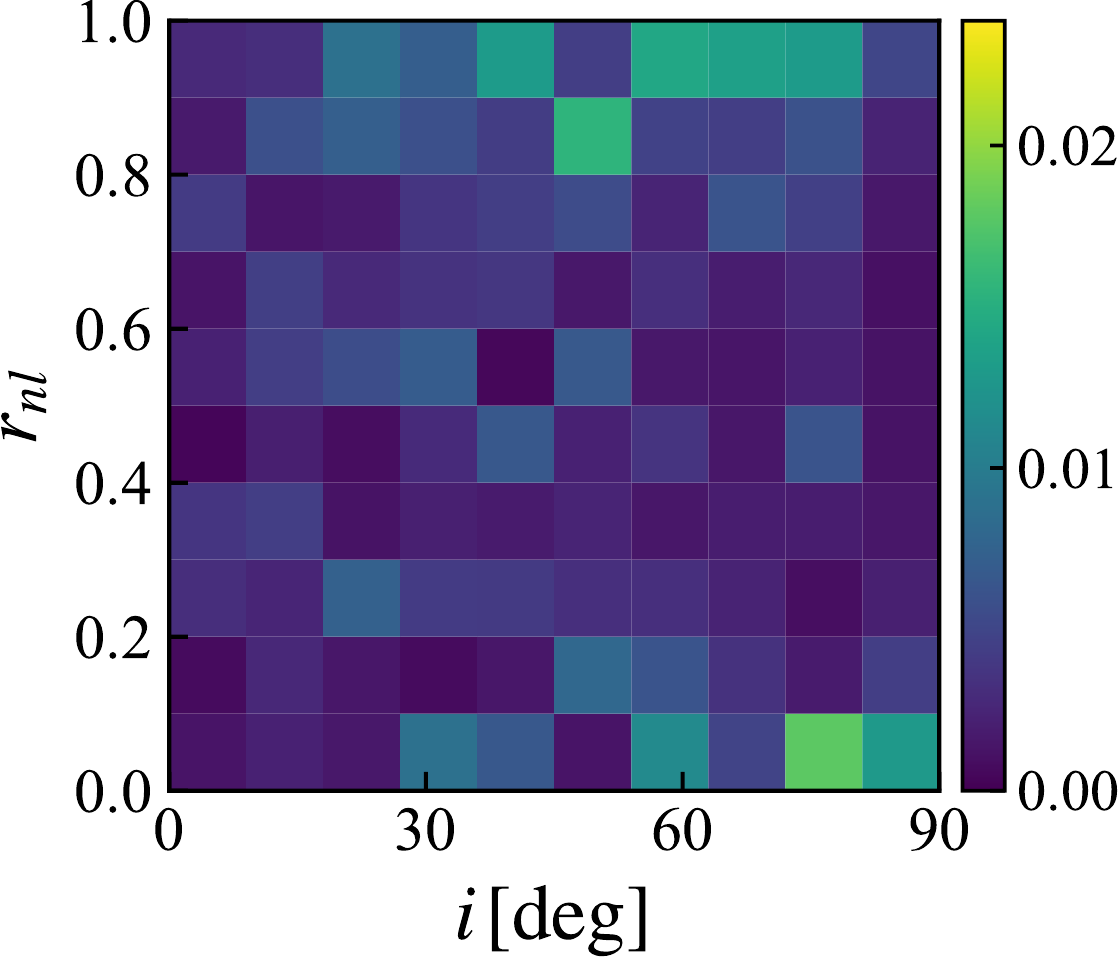} &
	\includegraphics[height=4.5cm]{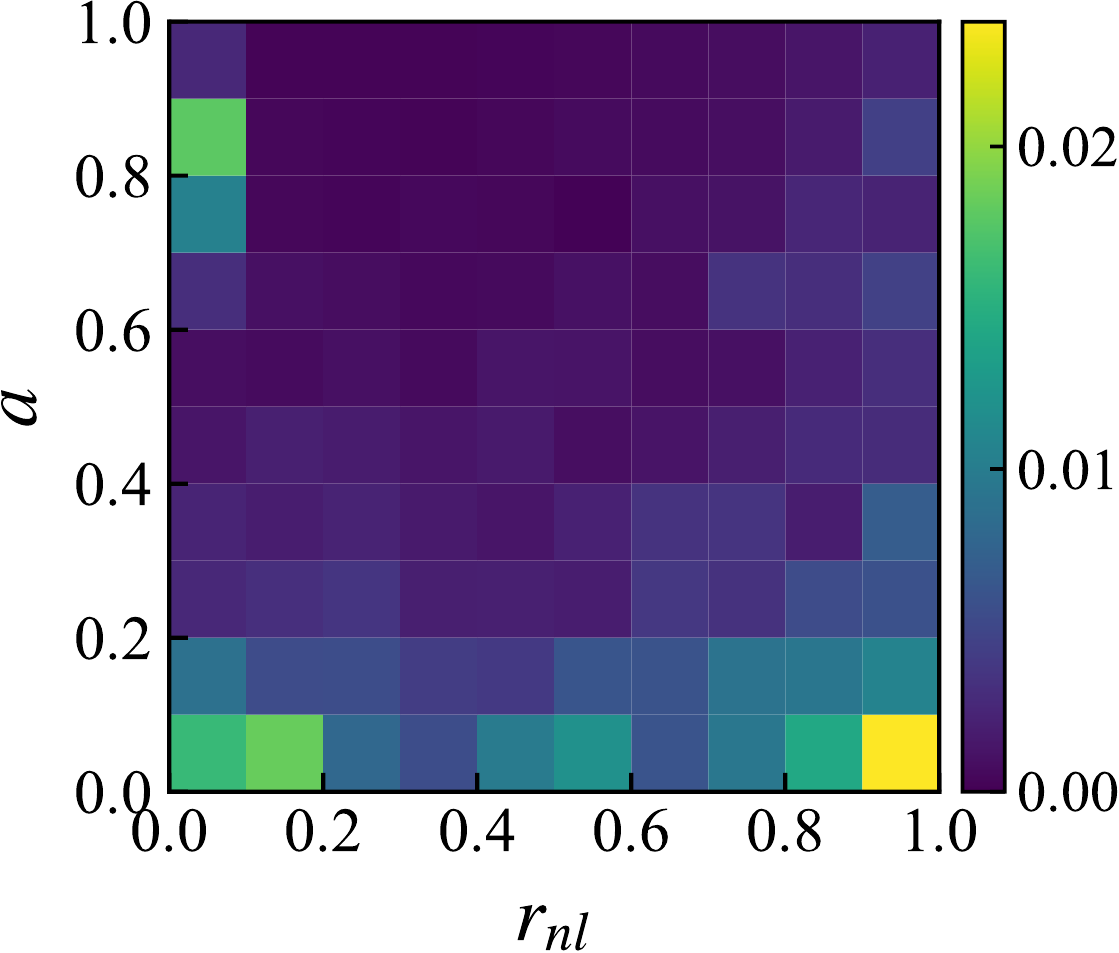} \\
	(a) $a$ error, $(a,i)$ &
	(b) $a$ error, $(i,r_{nl})$ &
	(c) $a$ error, $(r_{nl},a)$ \\
	\includegraphics[height=4.5cm]{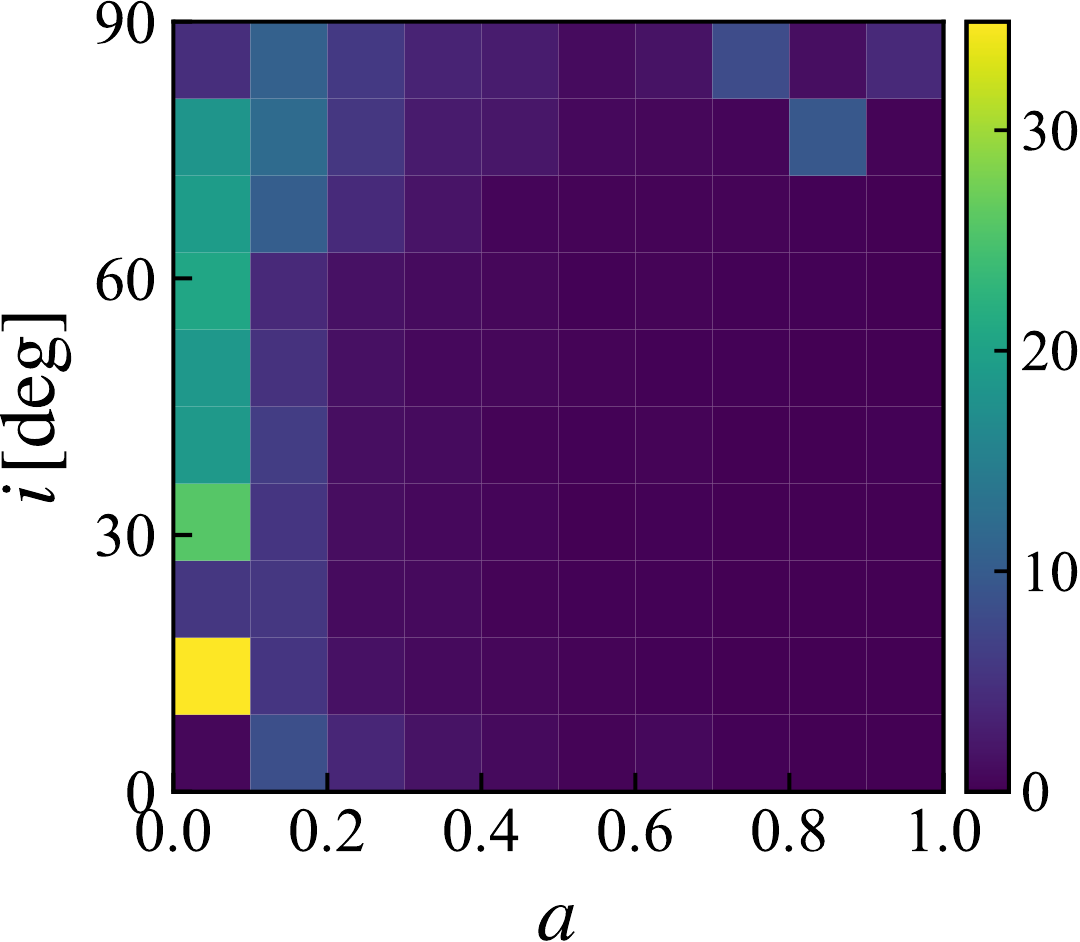} &
	\includegraphics[height=4.5cm]{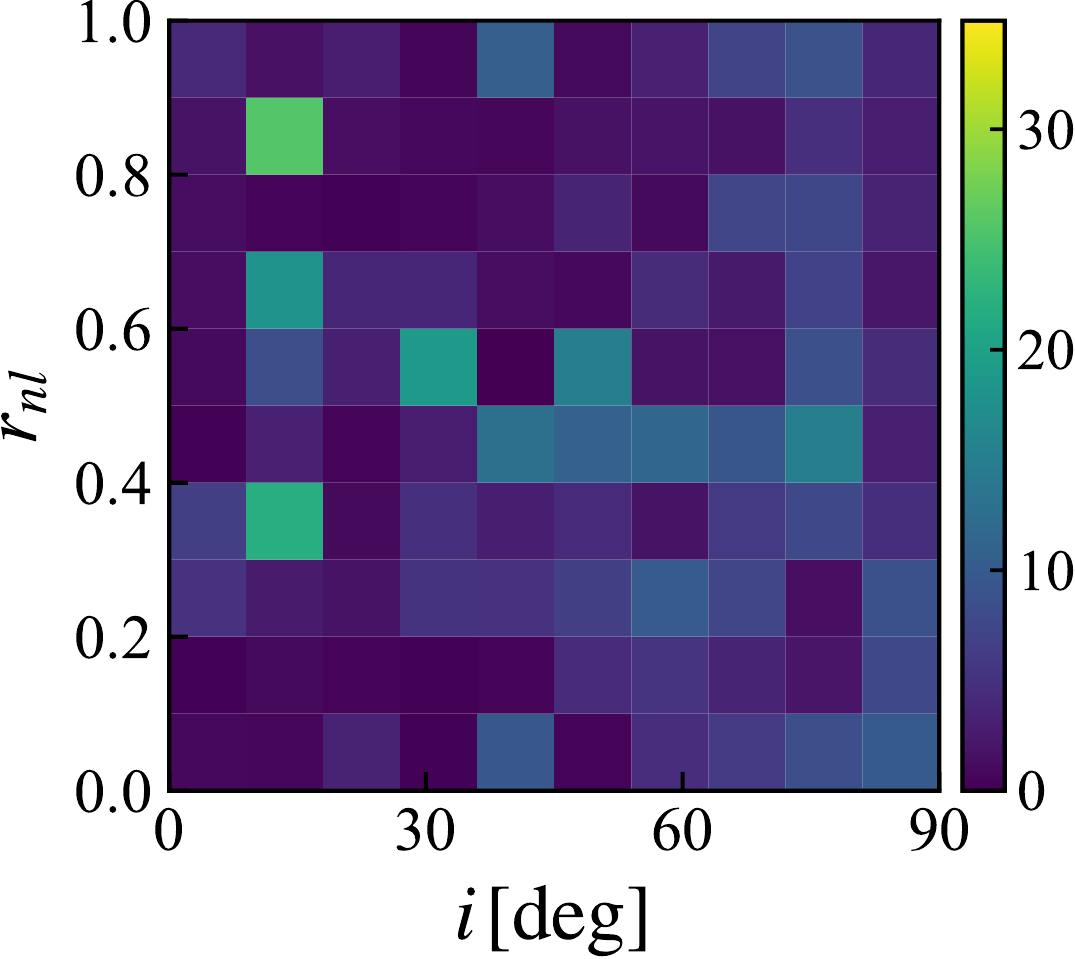} &
	\includegraphics[height=4.5cm]{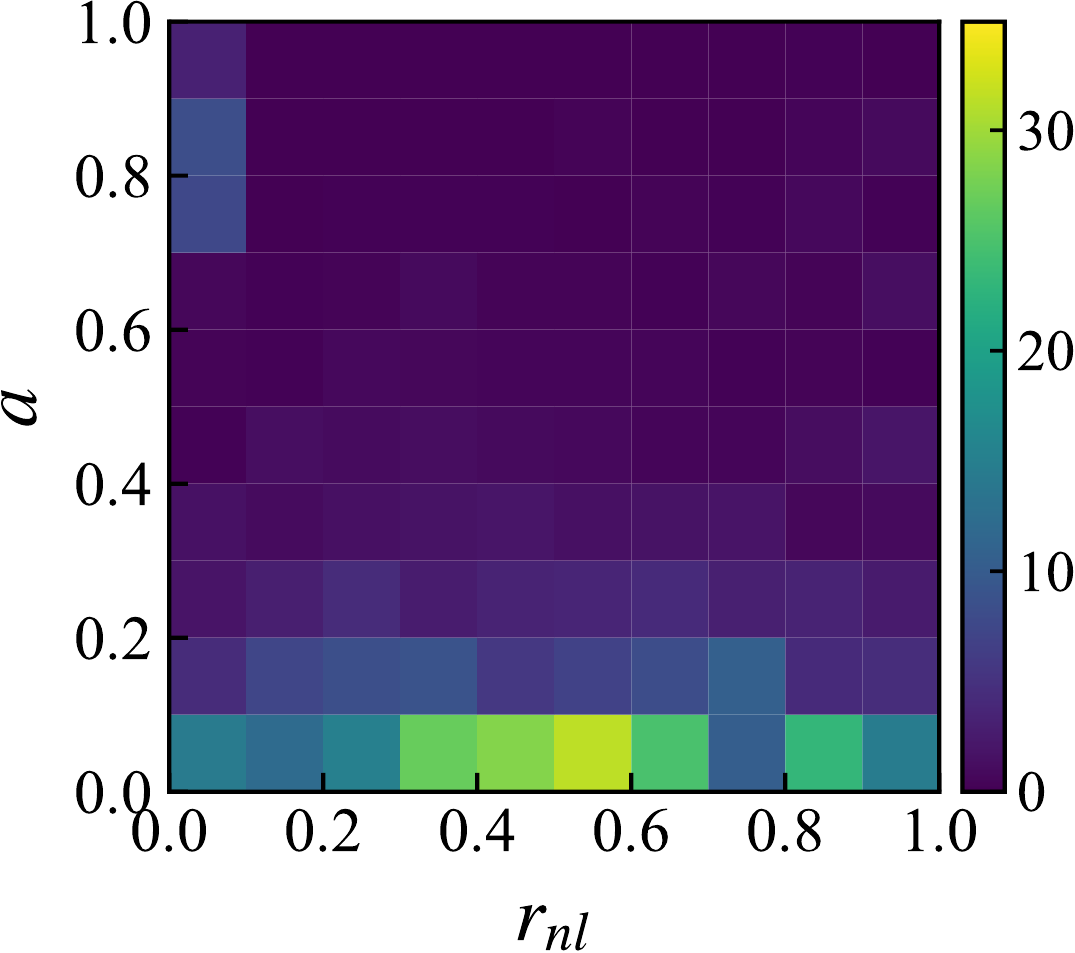} \\
	(d) $i$ error, $(a,i)$ &
	(e) $i$ error, $(i,r_{nl})$ &
	(f) $i$ error, $(r_{nl},a)$ \\
	\includegraphics[height=4.5cm]{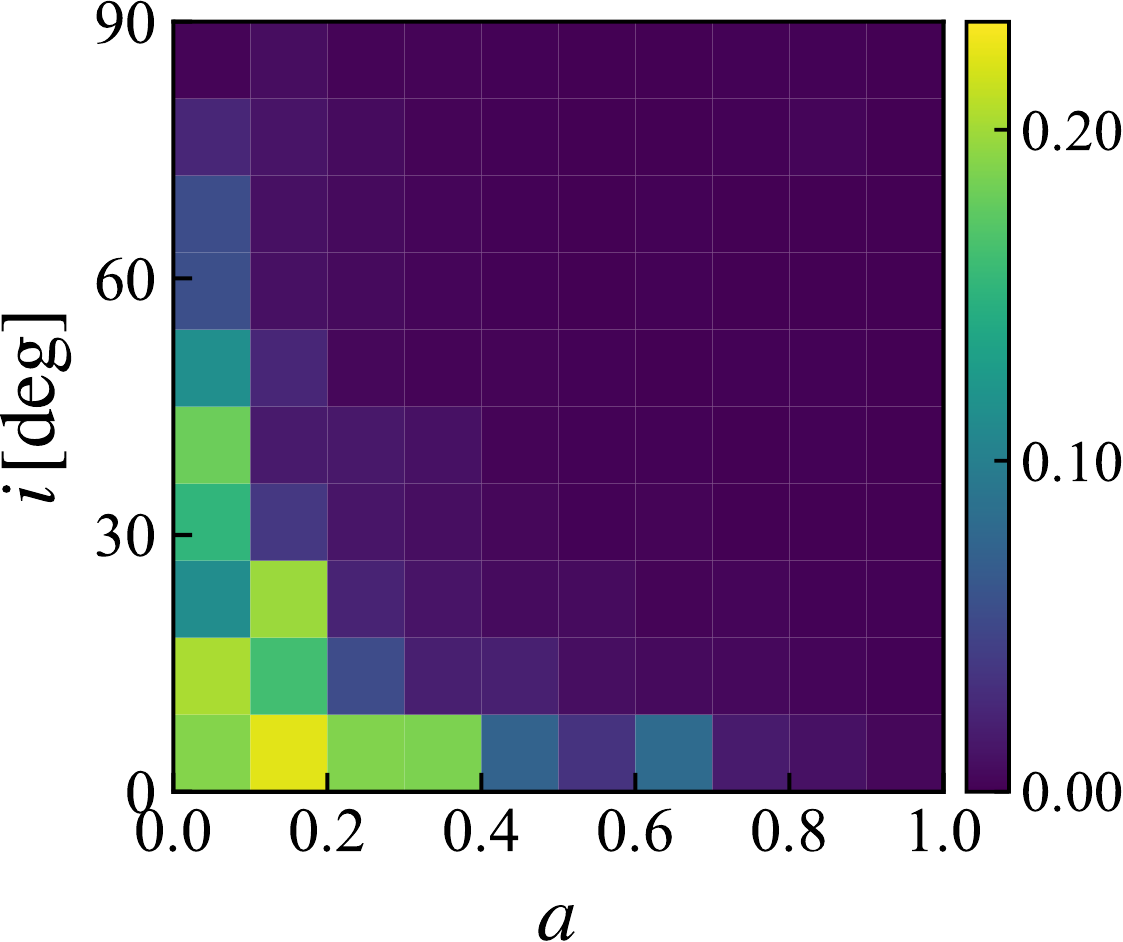} &
	\includegraphics[height=4.5cm]{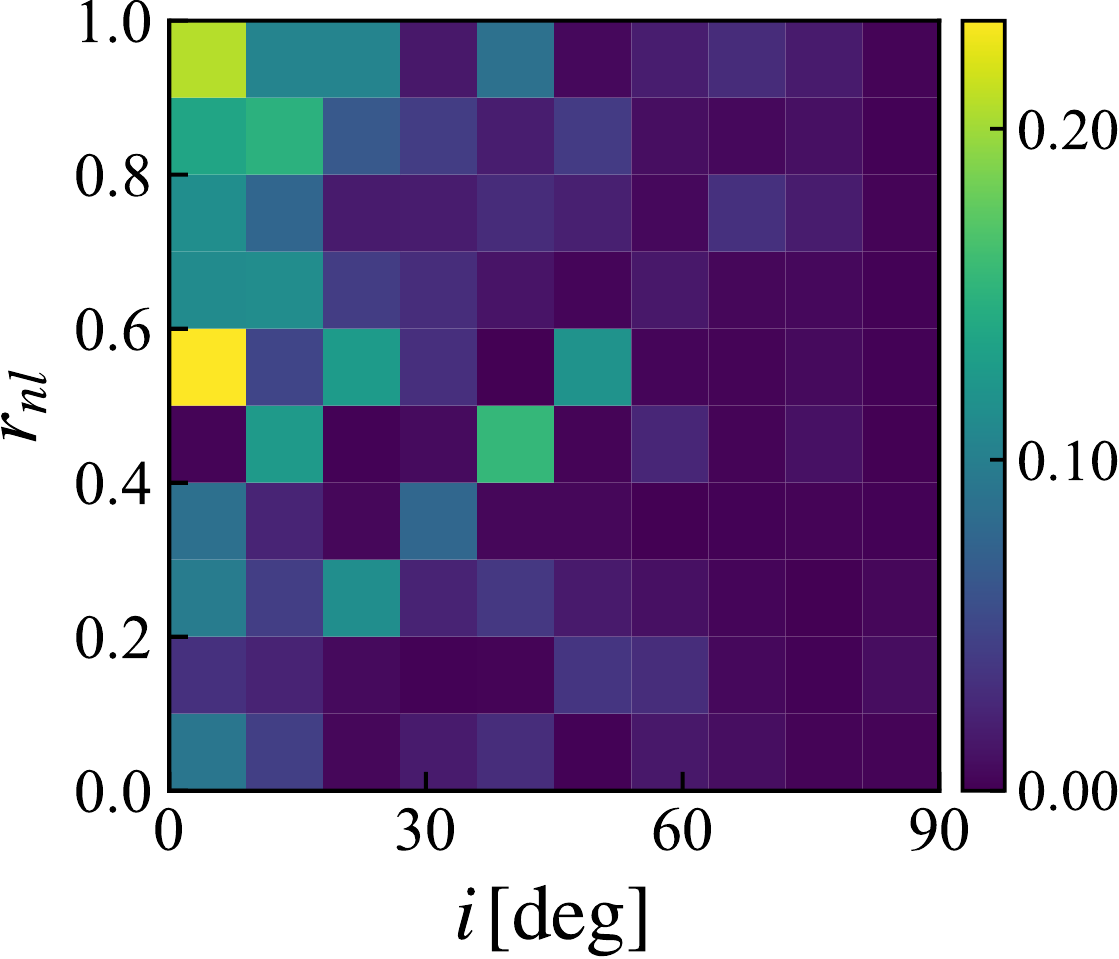} &
	\includegraphics[height=4.5cm]{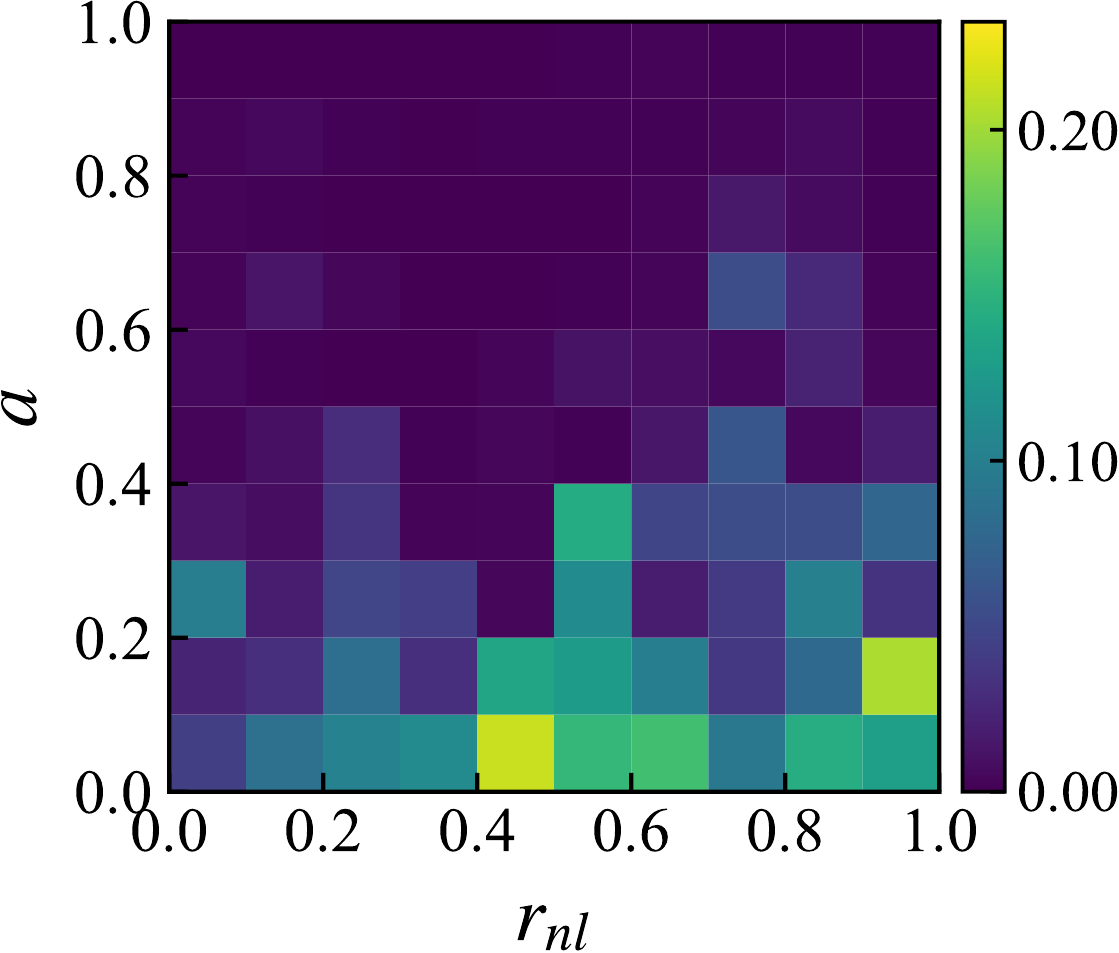} \\
	(g) $r_{nl}$ error, $(a,i)$ &
	(h) $r_{nl}$ error, $(i,r_{nl})$ &
	(i) $r_{nl}$ error, $(r_{nl},a)$ \\
	\end{tabular}
	\caption{\footnotesize{Local RMSE maps for the reconstructed parameters $a$, $i$, and $r_{nl}$
after artificial Gaussian perturbations with amplitude $\sigma=10^{-4}$
are added to the observables $(z_1,z_2,z_3)$.  Panels (a)--(c),
(d)--(f), and (g)--(i) show the local RMSE of $a$, $i$, and $r_{nl}$,
respectively.  In each row, the three columns correspond to the
parameter planes $(a,i)$, $(i,r_{nl})$, and $(r_{nl},a)$, respectively.
In each parameter plane, the evaluation samples are divided into
$10^2$ bins, and the local RMSE is computed for each bin.  The color
scale represents the local RMSE of the reconstructed parameter shown in
that row.}}
	\label{fig110}
\end{figure}
To examine the stability of the parameter-determination method against
uncertainties in the observables derived from the identified segments,
we also perform a perturbation test.  This test is not intended to model
the full observational pipeline, in which uncertainties may arise from
image reconstruction, segment identification, standardization, and the
estimation of Fourier coefficients.  Instead, it provides a diagnostic
of how perturbations in the observable values $(z_1,z_2,z_3)$ are
propagated to the reconstructed parameter values $(a,i,r_{nl})$.

We use the same $10^3$ evaluation samples as those used in
Sec.~\ref{sec:numerical}.  For each sample, artificial Gaussian
perturbations are added to the observables $(z_1,z_2,z_3)$.  The
perturbed observables are defined by
\begin{equation}
z_{j}^{p}
=
z_{j}
+
\sigma\,s_j\,\epsilon_{j},
\qquad
j=1,2,3 ,
\end{equation}
where $\epsilon_{j}$ are independent random numbers drawn from the
standard Gaussian distribution, and $s_j$ is the standard deviation of
$z_j$ over the $10^3$ evaluation samples.
The factor $s_j$ is used to normalize the artificial perturbation by the variation scale of each observable, because the three observables have different scales.
Thus, the dimensionless parameter $\sigma$ controls the perturbation amplitude relative to the
typical variation scale of each observable.

We then apply the parameter-determination method to the perturbed
observables $(z_1^{p},z_2^{p},z_3^{p})$ and obtain the corresponding
parameter values. The RMSE and MAE are computed by comparing these
values with the true parameter values.  In this way, we assess the
sensitivity of the method to uncertainties in the observables derived
from the identified segments.
Since the relation between actual
observational uncertainties and the standard deviation of $z_j$ is not
specified here, this test does not determine how small the observational
errors must be for the method to be accurate. Rather, it should be
regarded as a sensitivity test designed to examine how the reconstructed
parameter values $(a,i,r_{nl})$ respond to perturbations in the
observable values $(z_1,z_2,z_3)$.

Table~\ref{tab:reconstruction_errors_perturbed} summarizes the results
of the perturbation test.  For the large perturbation amplitude
$\sigma=10^{-1}$, the RMSE and MAE increase for all parameters, showing
that the parameter determination is strongly affected when the
observables are substantially perturbed.  As $\sigma$ decreases, the
RMSE and MAE also decrease.  For $\sigma\leq10^{-4}$, however, reducing
$\sigma$ further to $10^{-8}$ does not change the order of the RMSE and
MAE.  This indicates that the remaining errors are dominated by the
numerical reconstruction procedure rather than by the artificial
perturbations in the observables.  Across the perturbation amplitudes
shown in Table~\ref{tab:reconstruction_errors_perturbed}, the spin
parameter $a$ shows the smallest errors among the three parameters.

We evaluate the RMSE values by comparing them with the uninformative
and informative baselines introduced in Sec.~\ref{sec:numerical}.  For
$\sigma=10^{-1}$, the RMSE of $a$ is comparable with the
uninformative baseline, while the RMSE values of $i$ and $r_{nl}$
exceed the corresponding uninformative baselines.  These results
indicate that, when the observables are strongly perturbed, the
determination of all three parameters reaches an uninformative level in
terms of the RMSE.  For $\sigma\leq10^{-3}$, the RMSE of $a$ is below
the informative baseline, whereas the RMSE values of $i$ and $r_{nl}$
remain comparable with the corresponding uninformative baselines.

To examine where perturbations in the observables affect the
reconstruction in the parameter space, we compute local RMSE maps for
the representative case $\sigma=10^{-4}$.  Table~\ref{tab:reconstruction_errors_perturbed}
shows that, for $\sigma\leq10^{-4}$, further decreasing $\sigma$ does
not change the order of the RMSE and MAE, whereas larger perturbations
lead to a more pronounced increase in the errors.  Therefore,
$\sigma=10^{-4}$ is a suitable representative amplitude near the
transition where the effect of perturbations begins to become visible in
the reconstruction errors.

Figure~\ref{fig110} shows the resulting local RMSE maps for $a$, $i$,
and $r_{nl}$ with $\sigma=10^{-4}$.
As in Sec.~\ref{sec:numerical}, we use the 99th percentiles of the sample-wise absolute
errors as reference scales for interpreting these maps.
We regard the local RMSE of each parameter as relatively enhanced when it exceeds the corresponding 99th-percentile reference scale.
These values are $2.1350\times10^{-2}$, $28.704~[\mathrm{deg}]$, and
$3.2722\times10^{-1}$ for $a$, $i$, and $r_{nl}$, respectively.
The relatively enhanced local RMSE of the spin parameter $a$ remains localized.
The enhancement in the high-inclination and small-$r_{nl}$ region is similar to that seen
in the case without artificial perturbations in Fig.~\ref{fig100},
whereas the low-spin region becomes more pronounced in the perturbed
case.  For the inclination angle $i$, the enhanced local RMSE appears
mainly in the low-spin region, in contrast to the case without
artificial perturbations, where it appears mainly in the
high-inclination and small-$r_{nl}$ region.  For the parameter
$r_{nl}$, the enhanced local RMSE appears mainly in the low-spin and
low-inclination region.

In Fig.~\ref{fig110}, the regions with relatively enhanced local RMSE
appear more broadly in the $(i,r_{nl})$ plane than in the other two
parameter planes.  However, the maps that include the spin parameter $a$ show that these
enhancements are mainly associated with the low-spin region.  Since the
local RMSE in the $(i,r_{nl})$ plane is computed by grouping samples
with different values of $a$, the broad structure in this plane should
not be interpreted as a degradation over all spin values.  Rather, it
reflects the contribution from low-spin samples.

The regions with enhanced local RMSE in the perturbed case do not
necessarily coincide with those in the case without artificial
perturbations, because the two tests characterize different aspects of
the reconstruction error.  The unperturbed test mainly reflects
numerical interpolation and minimization errors, whereas the perturbed
test examines the sensitivity of the reconstruction to
perturbations in the observable values.  Therefore, the local RMSE can
be enhanced in regions where the observables are less sensitive to
changes in the parameters, even if the unperturbed reconstruction error
is small there.

The local RMSE enhancement in the low-spin region for $\sigma=10^{-4}$
can be understood from the geometry of the critical curve, as in the
case without artificial perturbations.  As discussed in
Sec.~\ref{sec:numerical}, the standardized segments in this region
depend only weakly on the parameters.  Without artificial perturbations,
the exact observable values may still select the correct parameter
values.  Once perturbations are added, however, this weak parameter
dependence can lead to relatively large changes in the reconstructed
parameters, making the enhancement more pronounced in the perturbed
maps.

To evaluate the relatively enhanced errors in Fig.~\ref{fig110}, we
compare the local RMSE values with the uninformative and informative baselines.
The local RMSE of $a$ remains below the informative baseline throughout the parameter space.
By contrast, the local RMSE of $r_{nl}$ is comparable with the
uninformative baseline in some localized regions. The local RMSE of
$i$ is comparable with, or exceeds, the uninformative baseline in some localized regions. Thus, for
$\sigma=10^{-4}$, the determination of $i$ and $r_{nl}$ by the present
method reaches an uninformative level in terms of the local RMSE in
those regions.

For the perturbed case with $\sigma=10^{-4}$, these results show that
the spin parameter $a$ exhibits the smallest local errors among the
three reconstructed parameters.  The parameter $r_{nl}$ is also
reconstructed with relatively small errors over most of the parameter
space, although localized enhancements appear in the low-spin and
low-inclination region.  By contrast, the inclination angle $i$ exhibits
larger local errors in the high-inclination and small-$r_{nl}$ region
and is more sensitive to perturbations in the observables, especially in
the low-spin region.

The enhanced local RMSE in the low-spin region in the perturbed case
also indicates that the loss of accuracy cannot be removed simply by
increasing the resolution of the parameter grid used in Supplemental
Material~\cite{SupplementalMaterial2}.  A denser grid may reduce
numerical interpolation and minimization errors.  However, in regions
where the observables are only weakly sensitive to changes in the
parameters, small perturbations in the observables are amplified by the
inverse mapping to the reconstructed parameters.  Therefore, increasing
the number of grid points alone is unlikely to fully resolve the loss of
accuracy in such regions.

This point is important for realistic applications.  The larger local
errors and stronger perturbation sensitivity of the inclination angle
$i$ indicate a practical limitation of determining $i$ using the
parameter-determination method.  In particular, it may be difficult to
determine $i$ accurately in the high-inclination and small-$r_{nl}$
region, and also in the low-spin region when the extracted observables
contain uncertainties, as expected in realistic observations due to
finite resolution, noise, and uncertainty in the identification of the
segment.  Additional information may therefore be needed, for example
independent constraints on the viewing geometry from the orientation of
the accretion flow or jet.

%-------------------------------------------%
\section{Conclusion}
\label{sec:con}
%-------------------------------------------%
In this paper, we have proposed a method for determining
black hole parameters from segments of the critical curve
in Kerr black hole images.
We restricted our analysis to non-extremal Kerr black holes
and introduced standardized segments of the critical curve, defined as connected curve segments in the upper
half-plane of the observer's screen whose endpoints are
separated by a distance of $2$.

We first reviewed null geodesics in Kerr spacetime and the role
of unstable spherical photon orbits in forming the critical curve.
Using Bardeen coordinates on the observer's screen, we
described segments of the critical curve as connected subsets
of the critical curve and clarified the differences among the
Schwarzschild, non-extremal Kerr, and extremal Kerr cases.

For non-extremal Kerr black holes, we showed that segments of the critical curves form a three-parameter family of embedded curve segments parametrized
by $(a,i,r_{nl})$.
We then introduced a representation of each standardized segment of the critical curve by expressing the curve in
polar coordinates through a radial function
$\rho(\varphi; a, i, r_{nl})$.
From the Fourier coefficients of this function up to seventh
order, we constructed a finite-dimensional description of the
shape of the standardized segment of the critical curve.

To extract informative observables from these Fourier data,
we applied principal component analysis and defined three
observables, $z_1$, $z_2$, and $z_3$, corresponding to the size,
the primary distortion, and the secondary distortion of the standardized segment of the critical curve.
Our numerical analysis of their isosurfaces indicates that the map
\begin{eqnarray}
(a,i,r_{nl}) \mapsto (z_1,z_2,z_3)
\end{eqnarray}
is injective over the considered domain.
Accordingly, once a standardized segment of the critical curve is extracted from observational data,
the parameters $(a,i,r_{nl})$ can be uniquely determined within
our framework.

We assessed the numerical performance and local error distribution of
the parameter-determination method without artificial perturbations in
the observables.  Using $10^3$ evaluation samples, the method gave RMSE
values of $2.4059\times10^{-3}$, $1.2511~[\mathrm{deg}]$, and
$9.1210\times10^{-3}$ for $a$, $i$, and $r_{nl}$, respectively.
These values are below the corresponding informative baselines.  The
local RMSE maps further showed that the local RMSE of $a$ remains below
the informative baseline throughout the parameter space, although it is
relatively enhanced in the high-inclination and small-$r_{nl}$ region.
By contrast, the local RMSE values of $i$ and $r_{nl}$ are comparable with the uninformative baseline in localized regions.
In particular, the local RMSE of $i$ is enhanced mainly in the high-inclination and small-$r_{nl}$ region,
whereas that of $r_{nl}$ is enhanced mainly in the low-spin and low-inclination region.
These localized enhancements are associated with regions where the standardized segments depend only weakly on the
parameters, so that different parameter values can produce similar standardized segments and hence similar observable values.

We also examined the sensitivity of the parameter-determination method
to artificial perturbations in the observables, treating them as a
diagnostic test rather than as a direct model of observational errors.
The RMSE values increased as the perturbation amplitude was increased.
For the largest perturbation considered, $\sigma=10^{-1}$, the RMSE of
$a$ was comparable with the uninformative baseline, while the RMSE
values of $i$ and $r_{nl}$ exceeded the corresponding uninformative
baselines. These results indicate that, when the observables are
strongly perturbed, the determination of all three parameters reaches
an uninformative level in terms of the RMSE.
For $\sigma\leq10^{-3}$, the RMSE of $a$ was below the informative
baseline, whereas the RMSE values of $i$ and $r_{nl}$ remained comparable with the corresponding uninformative baselines.
The local RMSE maps for $\sigma=10^{-4}$ further showed that the local
RMSE of $a$ remains below the informative baseline throughout the
parameter space. By contrast, the local RMSE of $r_{nl}$ is comparable with the uninformative baseline in the low-spin and
low-inclination region. The local RMSE of $i$ is comparable with, or
exceeds, the uninformative baseline in some localized regions, mainly in the low-spin region.
Thus, for $\sigma=10^{-4}$, the determination of $i$ and $r_{nl}$ by the present method reaches an uninformative level in terms of the local RMSE in those regions.

These results clarify both the usefulness and the limitation of the
method.  The determination of the spin parameter $a$ is the most stable
among the three parameters in our numerical tests.  However, the
determination of the inclination angle $i$ becomes less reliable in the
high-inclination and small-$r_{nl}$ region, and also in the low-spin
region.  This reduced reliability is not expected to be removed simply
by increasing the resolution of the parameter grid, because it is
associated with the weak parameter dependence of the observables rather
than only with numerical interpolation or minimization errors.

In realistic applications, the proposed method may provide relatively
robust constraints on the spin parameter $a$ from a segment of the
critical curve, whereas the inclination angle $i$ may require additional
information in unfavorable regions of the parameter space.  Such
information could include independent constraints on the viewing
geometry from the orientation of the accretion flow or jet.

Several extensions remain for future study.
First, a rigorous mathematical proof of injectivity would further strengthen the method.
Second, the robustness of the method against finite resolution, noise,
and uncertainties in segment identification should be investigated.
Third, it will be important to generalize the framework to more
realistic imaging conditions and to other black hole spacetimes,
including the extremal Kerr case and spacetimes with additional
parameters.

%-------------------------------------------%
\section*{Acknowledgements}
%-------------------------------------------%
This work was supported in part by JSPS KAKENHI Grant No. JP22K03623 (U.M.), No. JP20H05853 (T.H.) and No. JP24K07027 (T.H.).

\appendix

%------------------------------------------------------------%
\section{Reparameterization of Standardized Segments of Critical Curve}
\label{sec:re}
%------------------------------------------------------------%
In this appendix we show that standardized segment of the critical curve
features form embedded curves, and we describe their
reparametrization.

From Eqs.~(\ref{eq:spoell}) and (\ref{eq:alpha}) we obtain
\begin{eqnarray}
\frac{{\rm d}\alpha}{{\rm d}r_s}(r_s;a,i)
=
\frac{2\csc i\left[(r_s-1)^3+1-a^2\right]}
{a(r_s-1)^2}.
\label{eq:alphar}
\end{eqnarray}
For $r_s\in I(a,i)$ we have $r_s>1$.
Since $0<a<1$, it follows that $1-a^2>0$.
Hence the numerator of Eq.~\eqref{eq:alphar} is strictly positive,
and therefore
\begin{eqnarray}
\frac{{\rm d}\alpha}{{\rm d}r_s}(r_s;a,i)>0 .
\label{eq:dalphadr}
\end{eqnarray}

We define $r_{\min}(a,i)\coloneqq \min I(a,i)$.
Let $r_e(a,i)$ denote the value of $r_l$
for which the right endpoint of the standardized segment of the critical curve lies on the
$\alpha$ axis.
Thus $\alpha(r_s;a,i)$ is strictly monotone on
$[r_{\min}(a,i), r_e(a,i)]$,
and the parametrization $\gamma_{+}(r_s;a,i)$
is injective on this interval.

Moreover, since the critical curve is smooth~\cite{Paganini:2017qfo},
Eq.~\eqref{eq:dalphadr} implies that
\begin{eqnarray}
\frac{{\rm d}\gamma_+}{{\rm d}r_s}(r_s;a,i)
\neq
\boldsymbol{0}
\end{eqnarray}
on $I(a,i)$.
Hence $\gamma_+(r_s;a,i)$ provides a regular
parametrization of the upper half of the critical curve.

Consequently, its image
\begin{eqnarray}
\left\{
\gamma_+(r_s;a,i)
\,\middle|\,
r_s\in I(a,i)
\right\}
\end{eqnarray}
is an embedded curve; in particular, it is a Jordan arc.
Any subsegment of this embedded curve is again an embedded curve.
In particular, the segment
\begin{eqnarray}
\left\{
\gamma_+(r_s;a,i)
\,\middle|\,
r_s\in [ r_{\min}(a,i) , r_e(a,i)]
\right\}
\end{eqnarray}
is an embedded curve, since
$[ r_{\min}(a,i) , r_e(a,i)]\subset I(a,i)$.

We now reparametrize this curve as
\begin{eqnarray}
\left\{
\gamma_+(r_s;a,i)
\,\middle|\,
r_s\in[r_{\min}(a,i), r_e(a,i)]
\right\}
=
\left\{
\gamma_+(h_{a,i}(r_n);a,i)
\,\middle|\,
r_n\in[0,1]
\right\},
\end{eqnarray}
where $h_{a,i}$ is defined by
\begin{eqnarray}
h_{a,i}(r_n)
=
r_{\min}(a,i)
+
\bigl(r_e(a,i)-r_{\min}(a,i)\bigr)\,r_n.
\end{eqnarray}
The map $h_{a,i}$ is a diffeomorphism
\begin{eqnarray}
h_{a,i}:[0,1]\to[r_{\min}(a,i), r_e(a,i)].
\end{eqnarray}
Since reparametrization by a diffeomorphism preserves embeddedness,
the curve
\begin{eqnarray}
\left\{
\gamma_+(h_{a,i}(r_n);a,i)
\,\middle|\,
r_n\in[0,1]
\right\}
\end{eqnarray}
is also an embedded curve, and any of its subsegments is again an embedded curve.

We now reparametrize the standardized segment of the critical curve as
\begin{eqnarray}
\left\{
\gamma_+(h_{a,i}(r_n);a,i)
\,\middle|\,
r_n\in[r_{nl}, r_{nr}]
\right\},
\end{eqnarray}
where $r_{nl}, r_{nr}\in[0,1]$ with $r_{nl}<r_{nr}$.
The condition that the endpoints of the standardized segment of the critical curve are separated
by a fixed distance determines $r_{nr}$ uniquely
as a function of $r_{nl}$.

The standardized segments of the critical curves induce a smooth mapping
\begin{eqnarray}
C_p: P \times [0,1] \to \mathcal{A}(\mathbb{R}^2),
\end{eqnarray}
where $\mathcal{A}(\mathbb{R}^2)$ denotes the set of Jordan arcs
embedded in the observer's screen $\mathbb{R}^2$.
The mapping is defined by
\begin{eqnarray}
C_p(a,i,r_{nl})
=
\left\{
\gamma_+(h_{a,i}(r_n);a,i)
\,\middle|\,
r_n\in[r_{nl},\, r_{nr}]
\right\}.
\end{eqnarray}

Thus the family of standardized segments of the critical curves forms a three-parameter family of curves
on the observer's screen, parametrized by $(a,i,r_{nl})$,
while each individual standardized segment of the critical curve is parametrized by $r_n$.

%-------------------------------------------%

%-------------------------------------------%


\begin{thebibliography}{99}

%\cite{EventHorizonTelescope:2019dse}
\bibitem{EventHorizonTelescope:2019dse}
K.~Akiyama \textit{et al.} [Event Horizon Telescope],
%``First M87 Event Horizon Telescope Results. I. The Shadow of the Supermassive Black Hole,''
Astrophys. J. Lett. \textbf{875}, L1 (2019)
doi:10.3847/2041-8213/ab0ec7
[arXiv:1906.11238 [astro-ph.GA]].
%4745 citations counted in INSPIRE as of 19 Mar 2026

%\cite{EventHorizonTelescope:2022wkp}
\bibitem{EventHorizonTelescope:2022wkp}
K.~Akiyama \textit{et al.} [Event Horizon Telescope],
%``First Sagittarius A* Event Horizon Telescope Results. I. The Shadow of the Supermassive Black Hole in the Center of the Milky Way,''
Astrophys. J. Lett. \textbf{930}, no.2, L12 (2022)
doi:10.3847/2041-8213/ac6674
[arXiv:2311.08680 [astro-ph.HE]].
%2149 citations counted in INSPIRE as of 19 Mar 2026

%\cite{Bardeen:1973xx}
\bibitem{Bardeen:1973xx}
J.~M.~Bardeen,
in ``Black Holes,''
ed.\ C.\ DeWitt and B.\ DeWitt,
New York, Gordon \& Breach (1973).

%\cite{Chandrasekhar:1985kt}
\bibitem{Chandrasekhar:1985kt}
S.~Chandrasekhar,
``The mathematical theory of black holes'',
Oxford Univ.\ Press (1992).

%\cite{Hioki:2008zw}
\bibitem{Hioki:2008zw}
K.~Hioki and U.~Miyamoto,
%``Hidden symmetries, null geodesics, and photon capture in the Sen black hole,''
Phys. Rev. D \textbf{78}, 044007 (2008)
doi:10.1103/PhysRevD.78.044007
[arXiv:0805.3146 [gr-qc]].
%161 citations counted in INSPIRE as of 21 Mar 2026

%\cite{Hioki:2009na}
\bibitem{Hioki:2009na}
K.~Hioki and K.~i.~Maeda,
%``Measurement of the Kerr Spin Parameter by Observation of a Compact Object's Shadow,''
Phys. Rev. D \textbf{80}, 024042 (2009)
doi:10.1103/PhysRevD.80.024042
[arXiv:0904.3575 [astro-ph.HE]].
%521 citations counted in INSPIRE as of 18 Mar 2026

%\cite{Tsukamoto:2014tja}
\bibitem{Tsukamoto:2014tja}
N.~Tsukamoto, Z.~Li and C.~Bambi,
%``Constraining the spin and the deformation parameters from the black hole shadow,''
JCAP \textbf{06}, 043 (2014)
doi:10.1088/1475-7516/2014/06/043
[arXiv:1403.0371 [gr-qc]].
%217 citations counted in INSPIRE as of 18 Mar 2026

%\cite{Abdujabbarov:2015xqa}
\bibitem{Abdujabbarov:2015xqa}
A.~A.~Abdujabbarov, L.~Rezzolla and B.~J.~Ahmedov,
%``A coordinate-independent characterization of a black hole shadow,''
Mon. Not. Roy. Astron. Soc. \textbf{454}, no.3, 2423-2435 (2015)
doi:10.1093/mnras/stv2079
[arXiv:1503.09054 [gr-qc]].
%232 citations counted in INSPIRE as of 21 Mar 2026

%\cite{EventHorizonTelescope:2021dqv}
\bibitem{EventHorizonTelescope:2021dqv}
P.~Kocherlakota \textit{et al.} [Event Horizon Telescope],
%``Constraints on black-hole charges with the 2017 EHT observations of M87*,''
Phys. Rev. D \textbf{103}, no.10, 104047 (2021)
doi:10.1103/PhysRevD.103.104047
[arXiv:2105.09343 [gr-qc]].
%362 citations counted in INSPIRE as of 21 Mar 2026

%\cite{Hioki:2022mdg}
\bibitem{Hioki:2022mdg}
K.~Hioki and U.~Miyamoto,
%``Determining parameters of a spherical black hole with a thin accretion disk by observing its shadow,''
Phys. Rev. D \textbf{107}, no.4, 044042 (2023)
doi:10.1103/PhysRevD.107.044042
[arXiv:2210.02164 [gr-qc]].
%6 citations counted in INSPIRE as of 18 Mar 2026

%\cite{Hioki:2023ozd}
\bibitem{Hioki:2023ozd}
K.~Hioki and U.~Miyamoto,
%``Determining parameters of Kerr black holes at finite distance by shadow observation,''
Phys. Rev. D \textbf{109}, no.4, 044030 (2024)
doi:10.1103/PhysRevD.109.044030
[arXiv:2311.16802 [gr-qc]].
%3 citations counted in INSPIRE as of 21 Mar 2026

%\cite{Hioki:2024vta}
\bibitem{Hioki:2024vta}
K.~Hioki and U.~Miyamoto,
%``Determining parameters of Kerr{\textendash}Newman black holes by shadow observation from finite distance and spatial infinity,''
Class. Quant. Grav. \textbf{42}, no.12, 125009 (2025)
doi:10.1088/1361-6382/ade047
[arXiv:2411.08486 [gr-qc]].
%1 citations counted in INSPIRE as of 17 Jun 2026

%\cite{EventHorizonTelescope:2019ths}
\bibitem{EventHorizonTelescope:2019ths}
K.~Akiyama \textit{et al.} [Event Horizon Telescope],
%``First M87 Event Horizon Telescope Results. IV. Imaging the Central Supermassive Black Hole,''
Astrophys. J. Lett. \textbf{875}, no.1, L4 (2019)
doi:10.3847/2041-8213/ab0e85
[arXiv:1906.11241 [astro-ph.GA]].
%1618 citations counted in INSPIRE as of 19 Mar 2026

%\cite{Gralla:2019xty}
\bibitem{Gralla:2019xty}
S.~E.~Gralla, D.~E.~Holz and R.~M.~Wald,
%``Black Hole Shadows, Photon Rings, and Lensing Rings,''
Phys. Rev. D \textbf{100}, no.2, 024018 (2019)
doi:10.1103/PhysRevD.100.024018
[arXiv:1906.00873 [astro-ph.HE]].
%614 citations counted in INSPIRE as of 27 Dec 2025

%\cite{EventHorizonTelescope:2019pgp}
\bibitem{EventHorizonTelescope:2019pgp}
K.~Akiyama \textit{et al.} [Event Horizon Telescope],
%``First M87 Event Horizon Telescope Results. V. Physical Origin of the Asymmetric Ring,''
Astrophys. J. Lett. \textbf{875}, no.1, L5 (2019)
doi:10.3847/2041-8213/ab0f43
[arXiv:1906.11242 [astro-ph.GA]].
%1612 citations counted in INSPIRE as of 19 Mar 2026

%\cite{Narayan:1994xi}
\bibitem{Narayan:1994xi}
R.~Narayan and I.~s.~Yi,
%``Advection dominated accretion: A Selfsimilar solution,''
Astrophys. J. Lett. \textbf{428}, L13 (1994)
doi:10.1086/187381
[arXiv:astro-ph/9403052 [astro-ph]].
%1750 citations counted in INSPIRE as of 19 Mar 2026

%\cite{Yuan:2014gma}
\bibitem{Yuan:2014gma}
F.~Yuan and R.~Narayan,
%``Hot Accretion Flows Around Black Holes,''
Ann. Rev. Astron. Astrophys. \textbf{52}, 529-588 (2014)
doi:10.1146/annurev-astro-082812-141003
[arXiv:1401.0586 [astro-ph.HE]].
%1197 citations counted in INSPIRE as of 19 Mar 2026

%\cite{Dexter:2016cdk}
\bibitem{Dexter:2016cdk}
J.~Dexter,
%``A public code for general relativistic, polarised radiative transfer around spinning black holes,''
Mon. Not. Roy. Astron. Soc. \textbf{462}, no.1, 115-136 (2016)
doi:10.1093/mnras/stw1526
[arXiv:1602.03184 [astro-ph.HE]].
%171 citations counted in INSPIRE as of 19 Mar 2026

%\cite{Johnson:2019ljv}
\bibitem{Johnson:2019ljv}
M.~D.~Johnson, A.~Lupsasca, A.~Strominger, G.~N.~Wong, S.~Hadar, D.~Kapec, R.~Narayan, A.~Chael, C.~F.~Gammie and P.~Galison, \textit{et al.}
%``Universal interferometric signatures of a black hole{\textquoteright}s photon ring,''
Sci. Adv. \textbf{6}, no.12, eaaz1310 (2020)
doi:10.1126/sciadv.aaz1310
[arXiv:1907.04329 [astro-ph.IM]].
%351 citations counted in INSPIRE as of 18 Mar 2026

%\cite{Akiyama:2024msp}
\bibitem{Akiyama:2024msp}
K.~Akiyama, K.~Niinuma, K.~Hada, A.~Doi, Y.~Hagiwara, A.~E.~Higuchi, M.~Honma, T.~Kawashima, D.~Kolev and S.~Koyama, \textit{et al.}
%``The Japanese vision for the Black Hole Explorer mission,''
Proc. SPIE Int. Soc. Opt. Eng. \textbf{13092}, 130922E (2024)
doi:10.1117/12.3019968
[arXiv:2406.09516 [astro-ph.IM]].
%5 citations counted in INSPIRE as of 18 Mar 2026

%\cite{Johnson:2024ttr}
\bibitem{Johnson:2024ttr}
M.~D.~Johnson, K.~Akiyama, R.~Baturin, B.~Bilyeu, L.~Blackburn, D.~Boroson, A.~Cardenas-Avendano, A.~Chael, C.~k.~Chan and D.~Chang, \textit{et al.}
%``The Black Hole Explorer: motivation and vision,''
Proc. SPIE Int. Soc. Opt. Eng. \textbf{13092}, 130922D (2024)
doi:10.1117/12.3019835
[arXiv:2406.12917 [astro-ph.IM]].
%82 citations counted in INSPIRE as of 18 Mar 2026

%\cite{Lupsasca:2024xhq}
\bibitem{Lupsasca:2024xhq}
A.~Lupsasca, A.~C{\'a}rdenas-Avenda{\~n}o, D.~C.~M.~Palumbo, M.~D.~Johnson, S.~E.~Gralla, D.~P.~Marrone, P.~Galison, P.~Tiede and L.~Keeble,
%``The Black Hole Explorer: photon ring science, detection, and shape measurement,''
Proc. SPIE Int. Soc. Opt. Eng. \textbf{13092}, 130926Q (2024)
doi:10.1117/12.3019437
[arXiv:2406.09498 [gr-qc]].
%45 citations counted in INSPIRE as of 18 Mar 2026

\bibitem{Kerr} R.~P.~Kerr, Phys.\ Rev.\ Lett.\ {\bf 11}, 237 (1963).

%\cite{Carter:1968rr}
\bibitem{Carter:1968rr}
B.~Carter,
%``Global structure of the Kerr family of gravitational fields,''
Phys. Rev. \textbf{174}, 1559-1571 (1968)
doi:10.1103/PhysRev.174.1559
%2064 citations counted in INSPIRE as of 21 Mar 2026

%\cite{Chang:2020lmg}
\bibitem{Chang:2020lmg}
Z.~Chang and Q.~H.~Zhu,
%``Does the shape of the shadow of a black hole depend on motional status of an observer?,''
Phys. Rev. D \textbf{102}, no.4, 044012 (2020)
doi:10.1103/PhysRevD.102.044012
[arXiv:2006.00685 [gr-qc]].
%22 citations counted in INSPIRE as of 26 Mar 2023

%\cite{deVries:1999tiy}
\bibitem{deVries:1999tiy}
A.~de Vries,
%``The apparent shape of a rotating charged black hole, closed photon orbits and the bifurcation set $A_4$,''
Class. Quant. Grav. \textbf{17}, no.1, 123-144 (1999)
doi:10.1088/0264-9381/17/1/309
%203 citations counted in INSPIRE as of 06 Aug 2024

%\cite{Synge:1966okc}
\bibitem{Synge:1966okc}
J.~L.~Synge,
%``The Escape of Photons from Gravitationally Intense Stars,''
Mon. Not. Roy. Astron. Soc. \textbf{131}, no.3, 463-466 (1966)
doi:10.1093/mnras/131.3.463
%651 citations counted in INSPIRE as of 17 Mar 2026

%\cite{Gralla:2017ufe}
\bibitem{Gralla:2017ufe}
S.~E.~Gralla, A.~Lupsasca and A.~Strominger,
%``Observational Signature of High Spin at the Event Horizon Telescope,''
Mon. Not. Roy. Astron. Soc. \textbf{475}, no.3, 3829-3853 (2018)
doi:10.1093/mnras/sty039
[arXiv:1710.11112 [astro-ph.HE]].
%95 citations counted in INSPIRE as of 19 Mar 2026

%\cite{Paganini:2017qfo}
\bibitem{Paganini:2017qfo}
C.~F.~Paganini and M.~A.~Oancea,
%``Smoothness of the future and past trapped sets in Kerr\textendash{}Newman\textendash{}Taub-NUT spacetimes,''
Class. Quant. Grav. \textbf{35}, no.6, 067001 (2018)
doi:10.1088/1361-6382/aaaa5b
[arXiv:1710.02403 [gr-qc]].
%8 citations counted in INSPIRE as of 15 Nov 2024

\bibitem{Jolliffe}
Jolliffe, Ian T and Cadima, Jorge, ``Principal component analysis: a review and recent developments,'' Phil.\ Trans.\ R.\ Soc.\ A.\ {\bf 374}, 20150202 (2016).

\bibitem{SupplementalMaterial1}
See the Supplemental Material for table data on the transformation matrix $\bm{A}$.

\bibitem{SupplementalMaterial2}
See the Supplemental Material for table data on the dimensionless parameters, their corresponding Fourier coefficients, and principal components.

\end{thebibliography}
\end{document}